\documentclass[galaxies,review,accept,oneauthor,12pt,a4paper]{mdpi}

\usepackage{bm}
\usepackage{dcolumn}
\usepackage{amsmath}

\renewcommand \thesection{\arabic{section}}
\usepackage{epsfig}
\usepackage{soul}

\usepackage{amssymb,amsmath}
\usepackage{graphicx}
\usepackage{subfigure}
\makeatletter
\renewcommand{\@thesubfigure}{\normalsize(\textbf{\alph{subfigure}})}
\makeatother
\usepackage{array, dcolumn, multirow}
\usepackage{hyphenat}
\usepackage{enumitem,caption,booktabs}
\setitemize{parsep=0pt,itemsep=0pt,leftmargin=.4in}
\setenumerate{parsep=0pt,itemsep=0pt,leftmargin=.4in}

\lastpage{61}
\doinum{10.3390/galaxies2010022}
\pubvolume{2}
\pubyear{2014}
\history{Received: 6 November 2013; in revised form: 24 December 2013 / Accepted: 25 December 2013 /  Published: 17 January 2014}
\def \lleq {\lower0.9ex\hbox{ $\buildrel < \over \sim$} ~}
\def \ggeq {\lower0.9ex\hbox{ $\buildrel > \over \sim$} ~}

\newcommand{\ben}{\begin{eqnarray}}
\newcommand{\een}{\end{eqnarray}}

\def \beq  {\begin{equation}}
\def \eeq  {\end{equation}}
\def \ber  {\begin{eqnarray}}
\def \eer  {\end{eqnarray}}

\def \lleq {\lower0.9ex\hbox{ $\buildrel < \over \sim$} ~}
\def \ggeq {\lower0.9ex\hbox{ $\buildrel > \over \sim$} ~}

\def \beq  {\begin{equation}}
\def \eeq  {\end{equation}}
\def \ber  {\begin{eqnarray}}
\def \eer  {\end{eqnarray}}

\let\l=\left
\let\r=\right

\newcommand{\newc}{\newcommand}
\newc{\diag}{\mathop{\mathrm{diag}}}
\newc{\be}{\begin{equation}}
\newc{\ee}{\end{equation}}
\newc{\ba}{\begin{eqnarray}}
\newc{\ea}{\end{eqnarray}}
\newc{\bea}{\begin{eqnarray*}}
\newc{\eea}{\end{eqnarray*}}
\newc{\D}{\partial}
\newc{\ie}{{\it i.e.} }
\newc{\eg}{{\it e.g.} }
\newc{\etc}{{\it etc.} }
\newc{\etal}{{\it et al.}}
\newc{\lcdm}{$\Lambda$CDM}

\newc{\ra}{\rightarrow}
\newc{\lra}{\leftrightarrow}
\newc{\lsim}{\buildrel{<}\over{\sim}}
\newc{\gsim}{\buildrel{>}\over{\sim}}
\newc{\daa}{\left(\frac{\Delta \alpha}{\alpha}\right)}

\Title{Large Scale Cosmological Anomalies and Inhomogeneous \linebreak Dark Energy}

\Author{Leandros Perivolaropoulos}

\address[1]{%
Department of Physics, University of Ioannina, Ioannina 45110, Greece; E-Mail: leandros@uoi.gr; Tel.: +30-265-100-8632; Fax: +30-265-100-8698}

\corres{ }

\abstract{A wide range of large scale observations hint towards possible modifications on the standard cosmological model which is based on a homogeneous and isotropic universe with a small cosmological constant and matter. These observations, also known as ``cosmic anomalies'' include unexpected Cosmic Microwave Background perturbations on large angular scales, large dipolar peculiar velocity flows of galaxies (``bulk flows''), the measurement of inhomogenous values of the fine structure constant on cosmological scales (``alpha dipole'') and other effects. The presence of the observational anomalies could either be a large statistical fluctuation in the context of {\lcdm}  or it could indicate a non-trivial departure from the cosmological principle on Hubble scales. Such a departure is very much constrained by cosmological observations for matter. For dark energy however there are no significant observational constraints for Hubble scale inhomogeneities. In this brief review I discuss some of the theoretical models that can naturally lead to inhomogeneous dark energy, their observational constraints and their potential to explain the large scale cosmic anomalies.}


\begin{document}

%%%%%%%%%%%%%%%%%%%%%%%%%%%%%%%%%%%%%%%%%%

\section{Introduction}

The standard cosmological model (\lcdm) has been established through a wide range of cosmological observations \cite{Hamilton:2013dha} which challenged its validity during the past two decades. The model is based on the following assumptions: cosmological principle (isotropy, homogeneity, general relativity, cold dark matter and baryonic matter), flatness of space, the existence of a cosmological constant and gaussian scale invariant matter perturbations generated during inflation. Based on these assumptions, the model makes well defined predictions which are consistent with the vast majority of cosmological data. Some of these predictions include the following:

\begin{itemize}
\item
{\bf The Cosmic Microwave Background (CMB) Spectrum:} The angular power spectrum of CMB primordial perturbations \cite{Komatsu:2010fb} is in good agreement with the predictions of \lcdm.
However, a few issues related to the orientation and magnitude of low multipole moments (CMB anomalies) constitute remaining puzzles for the standard model \cite{Tegmark:2003ve,Copi:2010na,Bennett:2010jb,Schwarz:2004gk,Land:2005ad,Gruppuso:2010up,Sarkar:2010yj,Hanson:2009gu,Copi:2006tu}.

\item
{\bf The statistics of CMB perturrbations:} These statistics \cite{Smith:2009jr} are consistent with the prediction of gaussianity of the standard model \cite{Ade:2013ktc}.
\item
{\bf Accelerating expansion:} Cosmological observations using standard candles [Type Ia Supernovae (SnIa)]  \cite{Amanullah:2010vv} and standard rulers (Baryon Acoustic Oscillations)  \cite{Percival:2007yw} to map the recent accelerating expansion rate of the universe are consistent with the existence of a cosmological constant. No need has appeared for more complicated models based on dynamical dark energy or modified gravity, despite of the continuously improved data.  The likelihood of the cosmological \linebreak constant {\em vs}. more complicated homogeneous models has been continuously increasing during the \mbox {past decade \cite{lpsniarev}}.
\item
{\bf Large Scale Structure:} Observations of large scale structure are in good agreement with \lcdm~\cite{Nesseris:2007pa} (basic statistics of galaxies  \cite{TrujilloGomez:2010yh}, halo power spectrum \cite{Reid:2009xm}).
\end{itemize}

Despite of the above major successes the standard model is challenged by a few puzzling large scale cosmological observations \cite{Perivolaropoulos:2008ud,Yang:2009ae} which may hint towards required modifications of the model. Some of these challenges of {\lcdm} may be summarized as follows:
\begin{enumerate}
\item
{\bf Power Asymmetry of CMB perturbation maps:}  A hemispherical power asymmetry in the cosmic microwave background (CMB) on various different angular scales (multipole ranges) has been detected \cite{Ade:2013nlj,Eriksen:2007pc,Paci:2010wp,Hoftuft:2009rq}. The power in all  multipole ranges is consistently found to be significantly higher in the approximate direction towards Galactic longitude and latitude ($l = 237^\circ, b = -20^\circ$) than in the opposite direction. A more recent study of the WMAP9 data has found a hemispherical directional dependence of {\lcdm} cosmological parameters which is maximized at the direction ($l = 227^\circ, b = -27^\circ$) at the multipole range $2$--$600$ and is statistically significant at the $3.4$$\sigma$ level \cite{Axelsson:2013mva}. A related asymmetry, is the Maximum Temperature Asymmetry (MTA) defined as the maximized temperature difference between opposite pixels in the sky which shows significant alignment with other apparently unrelated asymmetries \cite{Mariano:2012ia}.
\item
{\bf Large Scale Velocity Flows:}  Recent studies have indicated the existence of dipole velocity flows on scales of 100 h$^{-1}$ Mpc \cite{Watkins:2008hf,Feldman:2009es} with magnitude about $400$ km/s using a combination of peculiar velocity surveys, with direction towards  $l \simeq 282^\circ$, $b\simeq 6^\circ$. Other studies \cite{Kashlinsky:2008ut} using the kinematic S-Z effect have found bulk flows on much larger scales ($O$(1 Gpc)) with magnitude \mbox {$600$--$1000$ km/s} towards a similar direction. These results are inconsistent with the predictions of {\lcdm} at a level of $99\%$. These studies however have been challenged by other authors which do not confirm these results but find peculiar velocities on these scales consistent with \lcdm~\cite{Lavaux:2008th}. Even though these studies agree with the direction of the observed flow they disagree on the magnitude and errorbar of the measured velocities.

\item
{\bf Alignment of low multipoles in the CMB angular power spectrum:} The normals to the octopole and quadrupole planes are aligned with
the direction of the cosmological dipole at a level inconsistent with Gaussian random, statistically isotropic skies at 99.7\% \cite{Copi:2010na}. This inconsistency has been reduced by the recent Planck results to a level of about 98$\%$ \cite{Ade:2013nlj} (the exact level varies slightly depending on the foreground filtering method).
\item
{\bf Large scale alignment in the QSO optical polarization data:} Quasar polarization vectors are not randomly oriented over the sky with a probability often in excess of 99.9\%. The alignment effect seems to be prominent along a particular axis in the direction $(l,b)=(267^\circ, 69^\circ)$ \cite{Hutsemekers:2001xm,Hutsemekers:2005iz,Hutsemekers:2005mods}.
\item
{\bf Anisotropy in Accelerating Expansion Rate:} Recent studies of the accelerating cosmic expansion rate using SnIa as standard candles have indicated that an anisotropic expansion rate fit by a dipole provides a better fit to the data than an isotropic expansion rate at the $2$$\sigma$ level \cite{Mariano:2012wx} Interestingly, this dark energy dipole is abnormally aligned \cite{Mariano:2012wx} with the fine structure constant dipole discussed below.
\item
{\bf Spatial dependence of the value of the fine structure constant $\alpha$:} A spatial cosmic variation of the fine structure constant has been recently identified on redshifts up to $z\simeq 4.2$ by analyzing the absorption spectra of quasars. This anisotropy analysis of the fine structure constant $\alpha$ \cite{King:2012id,King:2012id2} is based on a large
sample of quasar absorption-line spectra (295 spectra) obtained using UVES (the Ultraviolet and Visual
Echelle Spectrograph) on the VLT (Very Large Telescope) in Chile and also previous observations at the Keck Observatory in Hawaii. An apparent variation of $\alpha$ across
the sky was found. It was shown to be well fit  by an angular dipole model $\daa=A \cos\theta + B$ where $\theta$ is the angle with respect to a preferred axis and $A,B$ are the dipole magnitude and an isotropic monopole term. The dipole axis was found to point in the direction
$(l,b)= (331^\circ, -14^\circ)$ and the dipole amplitude $A$ was found to be $A = (0.97\pm 0.21)\times 10^{-5}$. The statistical significance over an isotropic model was found to be at the $4.1$$\sigma$ level.
\item
{\bf Large Quasar Group:} An elongated structure of quasars with long dimension about $1240$ Mpc and mean redshift ${\bar z} = 1.27$ has recently been discovered \cite{Clowes:2012pn}. This structure is a factor of about three larger than the previously known largest structure (Sloan Great Wall (${\bar z}= 0.073$ and comoving size $450$ Mpc \cite{Gott:2003pf}) and appears to be inconsistent with the cosmological principle in the context of the standard {\lcdm} model in the sense that it is much larger than the scale of homogeneity in the context of \lcdm~(260--370 Mpc \cite{Yadav:2010cc}).
\end{enumerate}

\begin{figure}[H]
\centering
\includegraphics[scale=0.8]{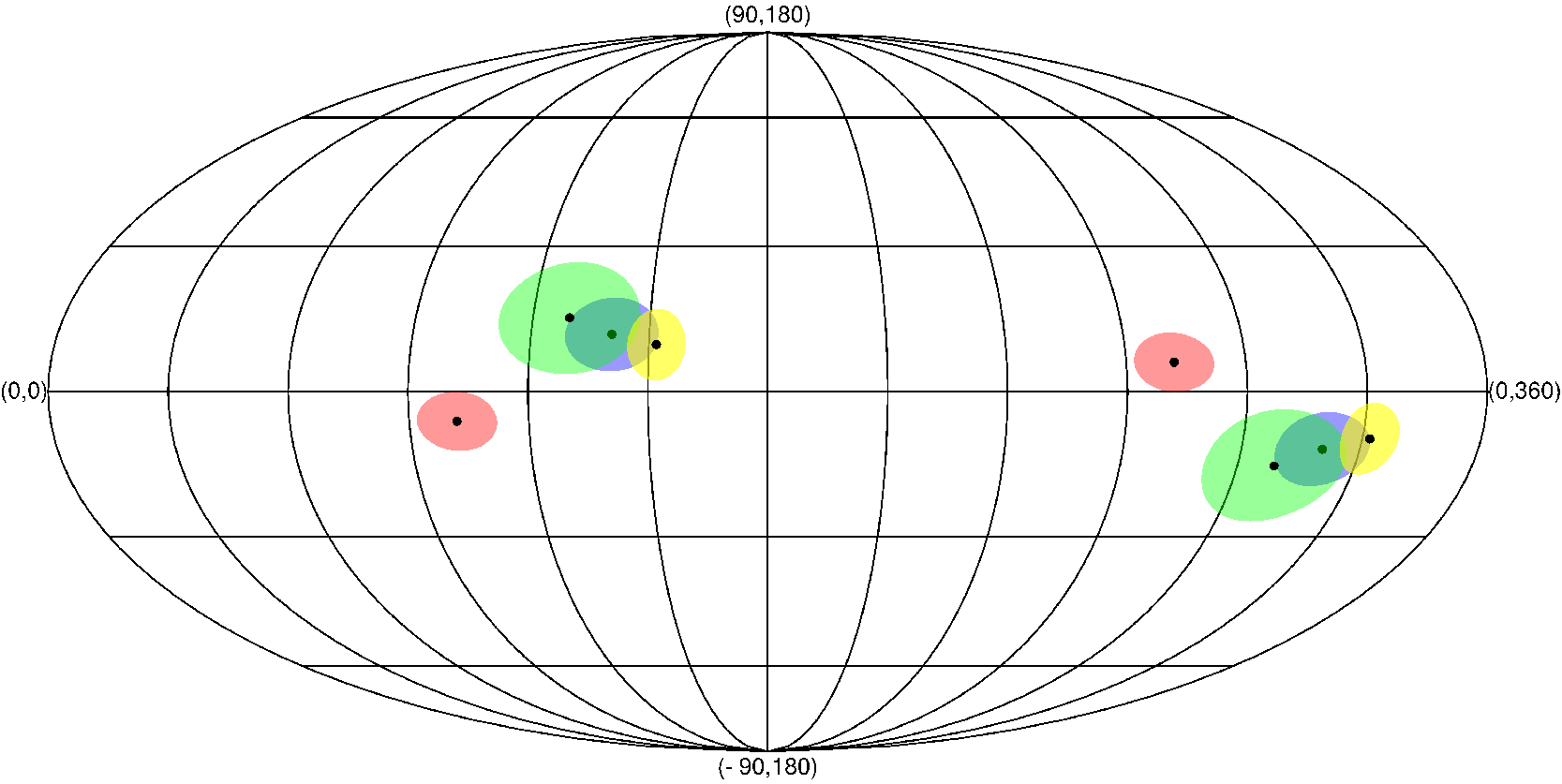}
\caption{Directions in galactic coordinates for the $\alpha$ (blue)
  and Dark Energy (green) dipoles, for the Dark Flow direction (red)
  and for the direction of Maximum Temperature Asymmetry (MTA) in the 7
  years ILC CMB map degraded to $N_{side}=8$ (yellow). The opposite corresponding directions are also shown (from  \cite{Mariano:2012ia}). }
\label{fig:error_blobs}
\end{figure}

The above seven large scale puzzles are large scale effects (1 Gpc or larger) and seem to be related to violation of the cosmological principle. In fact preferred cosmological directions may be associated with most of them. As discussed in more detail in Section 3, some of these directions appear to be abnormally close to each other. These directions are shown in Figure \ref{fig:error_blobs} \cite{Mariano:2012ia}. This may be an indication for a common physical origin at least for some of the above anomalies.

These anomalies may be simply large statistical fluctuations that manifest themselves duo to a posteriori selection of particular statistical tests. In view of their diverse nature and their high statistical significance, is not likely that they are all of this nature. In view of the possibility that there is a physical mechanism that leads to their existence, it is of interest to identify classes of physical models that could predict the existence of some or all of the above cosmic anomalies.

The origin of the possible violation of the cosmological principle could be either geometric or energetic. \scalebox{.95}{A geometric origin would correspond to inherent properties of the cosmic metric (e.g., non-trivial} topology on large scales). An energetic origin would correspond to non-trivial properies of the energy momentum tensor of the contents of the universe (e.g., anisotropic dark energy equation of state).

Models based on geometric violation of the cosmological principle include the following:
\begin{itemize}
\item
A multiply connected non-trivial cosmic topology \cite{Luminet:2008ew,Bielewicz:2008ga}.
\item
A rotating universe coupled to an anisotropic scalar field \cite{Carneiro:2001fz}.
\item
Non-commutative geometry \cite{Akofor:2007fv}.
\item
A fundamental anisotropic curvature \cite{Koivisto:2010dr}.
\end{itemize}

Models based on energetic violation of the cosmological principle include:
\begin{itemize}
\item
Anisotropic dark energy equation of state \cite{Zumalacarregui:2010wj,Koivisto:2005mm,Battye:2009ze} due perhaps to the existence of vector \mbox {fields \cite{ArmendarizPicon:2004pm,EspositoFarese:2009aj}}.
\item
Dark matter perturbations on a few Gpc scale \cite{Rodrigues:2007ny,Jimenez:2008vs}. For example an off center observer in a 1 Gpc void would experience the existence of a preferred cosmological axis through the \linebreak Lema\^itre-Tolman-Bondi metric \cite{Alexander:2007xx,GarciaBellido:2008nz,Biswas:2010xm,Dunsby:2010ts,Garfinkle:2009uf}. Within this framework, an additional dark energy component is not needed to secure consistency with the cosmological data that indicate accelerating expansion. The basic idea is that the increased expansion rate occurs locally in space rather than at recent cosmological times, a fact that can be achieved by assuming a locally-reduced matter energy density~\cite{AndrzejKrasinski:1997zz,Bolejko:2011jc}. Thus, the observer is placed close to the center of a giant void with dimensions of a few Gpc~\cite{Alnes:2005rw}. Even though this approach is free of dark energy, it is by no means free of fine tuning. Apart from the unnatural assumption of giant-size {\rm \,Gpc} voids, which are very unlikely to be produced in any cosmology, these models require the observer to be placed within a very small volume at the center of the void (about $10^{-6}$ of the total volume of the void). A slightly off-center observer, however, will naturally experience a preferred cosmological direction (towards the center of the void), which may help to resolve some of the observational puzzles of {\lcdm} discussed above.

Such matter perturbations could be induced by statistically anisotropic or non-gaussian primordial perturbations \cite{ArmendarizPicon:2007nr,Pullen:2007tu,Ackerman:2007nb,ValenzuelaToledo:2010cs}. For example, inflationary perturbations induced by vector fields \cite{Dimopoulos:2008yv,Yokoyama:2008xw,Golovnev:2009ks,Bartolo:2009pa}. Note however that inflationary models with vector fields usually suffer from instabilities due to the existence of ghosts \cite{ghosts,ghosts1,ghosts2}.
\item
Large scale electromagnetic fields \cite{Thorsrud:2012mu}. For example, the existence of a large scale primordial magnetic field \cite{Kahniashvili:2008sh,Barrow:1997mj,Campanelli:2009tk}. Evidence for such a magnetic field has recently been found in CMB \mbox {maps \cite{Kim:2009gi}}.
\item
Dark energy perturbations on scales comparable to the horizon. Even though the sound speed for dark energy is close to unity implying that it can not cluster on scales much smaller than the horizon, it can still produce observable effects due to clustering on horizon scales \cite{Bean:2003fb,BuenoSanchez:2011zz}. \linebreak In addition the possible recent formation of topological defects with Hubble scale core (topological quintessence) could also behave as inhomogeneous dark energy \cite{BuenoSanchez:2011wr}.
\end{itemize}

Models based on matter underdensities (voids) on scales larger than 1 Gpc are severely constrained by observations \cite{Marra:2011ct,Zumalacarregui:2012pq,GarciaBellido:2008nz,Caldwell:2007yu,GarciaBellido:2008gd} and do not reduce to {\lcdm}  for any value of their parameters. For radii as large as the horizon they reduce to homogeneous open CDM matter dominated universe and for small radii they reduce to flat CDM models without a cosmological constant. Both of these classes are ruled out by observations. Even for radii of order a few  Gpc these models are faced by severe problems that can be summarized as follows:
\begin{itemize}
\item
There is no simple physical mechanism to generate such large voids \cite{Ricciardelli:2013kxa}.
\item
As discussed above, off center observers experience a CMB dipole that is in excess of the observed one. Thus our location is constrained to practically coincide with the center of the void (fine tuning) to within about a fraction of a percent in radious \cite{Alnes:2006pf,Grande:2011hm}.
\item
The fit for the expansion rate as a function of redshift is worse than {\lcdm} requires significant fine tuning of the void profile to obtain a comparable fit with {\lcdm} \cite{Grande:2011hm}.
\item
These models predict significant peculiar velocities for distant galaxies within the void. Such velocities could have been detected in CMB maps through the kinematic S-Z effects. The fact that such large radial velocities are not observed imposes the most severe class of constraints in this class of models \cite{Caldwell:2007yu}.
\end{itemize}

In contrast to matter void models, spherically inhomogeneous dark energy models are more natural and reduce to {\lcdm} when the inhomogeneity radious exceeds the horizon scale. They constitute one of the two generic generalizations of \lcdm. Instead of breaking time translation invariance of the cosmological constant (as is the case for quintessence models) inhomogeneous dark energy models break space translation invariance of $\Lambda$. This corresponds to addressing the coincidence problem of {\lcdm} not as ``why now?'' question but as a ``why here?'' question. The motivation for considering this class of inhomogeneous dark energy models may be summarized as follows:
\begin{itemize}
\item
It is a new generic generalization of {\lcdm} including {\lcdm} as a special case.
\item
It naturally violates the cosmological principle on large cosmic scales and predicts a preferred axis for off center observers. Thus it has the potential to address at least some of the above discussed cosmic anomalies.
\item
There is a well defined physical mechanism that can give rise to this type on dark energy inhomogeneities. This mechanism will be discussed in some detail in Section 3. It is based on applying the principles of {\it topological inflation}~\cite{Vilenkin:1994pv} to the case of late-time acceleration. According to the idea of topological inflation, the false vacuum energy of the core of a topological defect can give rise to accelerating expansion if the core size reaches the Hubble scale when gravity starts dominating the dynamics. Thus, for example, a recently formed global monopole with appropriate scale of symmetry breaking and coupling could naturally produce a Hubble-scale, spherically symmetric, isocurvature dark energy overdensity. By analogy with topological inflation, this mechanism may be called {\it topological quintessence}.
\end{itemize}

In this review we focus on models which can violate the cosmological principle via energy momentum tensor effects (energetic origin violations). Since Hubble scale matter inhomogeneities are severely constrained we consider mainly dark energy Hubble scale inhomogeneities and/or anisotropies which appear to be a more natural source of a mild violation of the cosmological principle.

The structure of this review is the following: In the next section we present a collective analysis of the above discussed cosmic anomalies. After describing in some detail the current observational status of  each one of these anomalies we discuss the statistical significance of its incompatibility with large scale homogeneity and isotropy. In Section 3 we describe theoretical models that predict the existence of anisotropic or inhomogeneous dark energy and the potential of these models to explain some of the cosmic anomalies. We discuss two classes of models: Anisotropic dark energy models described by a Bianchi I cosmic metric with anisotropic dark energy equation of state and spherical inhomogeneous dark energy models described by a generalization of the Lema\^itre-Tolman-Bondi metric. A special case of the later class of models is ``topological quintessence'' where a well defined physical mechanism generates the spherical dark energy inhomogeneity. Finally in Section 4 we conclude and summarize our results. We also present an outlook discussing remaining open questions and possible future directions of research related to this subject.

\section{A Review of Cosmic Anomalies}

In this section we discuss in some detail the cosmological observations that hint towards a possible violation of the large scale homogeneity and isotropy of the universe.

\subsection{Power Asymmetry of CMB Perturbation Maps}

As discussed in the Introduction, the power spectrum of CMB fluctuations as measured by Wilkinson Microwave Anisotropy Probe (WMAP) and Planck is not isotropic. A hemispherical asymmetry has been pointed out in several \mbox {studies \cite{Ade:2013nlj,Komatsu:2010fb,Axelsson:2013mva,ghosts2,Hansen:2008ym,Eriksen:2003db,Park:2003qd}} using both the WMAP \cite{Komatsu:2010fb} and the \mbox {Planck data \cite{Ade:2013nlj}} at several multipole ranges. The persistense of this asymmetry in different multipole ranges along with the good multifrequency component separation (especially in the Planck data) make it unlikely that the asymmetry is due to foregrounds. The observed asymmetry is modeled as a power spectrum dipolar modulation of an otherwise statistically isotropic sky:
\be
\Delta T({\hat n})=(1+A {\hat p}\cdot{\hat n}) \Delta T_{iso}({\hat n}) \label{dipolanis}
\ee
where ${\hat p}$ and ${\hat n}$ are unit vectors towards the dipole and the sky directions respectively and $\Delta T_{iso}({\hat n})$ is a statistically isotropic temperature fluctuation in the direction $\hat n$. The best fit dipole is found to have direction $(l,b)=(227,-27)$ and amplitude $A=0.072\pm 0.022$ for the multipole range $l<64$ \linebreak ($k <$ 0.035 Mpc$^{-1}$) \cite{Ade:2013nlj}. The derived amplitude is inconsistent with isotropy ($A=0$) at the $3.4$$\sigma$ level. If this dipole modulation of the angular power spectrum is not a statistical fluctuation and has a physical cause, then this cause should be such as to change the amplitude of {\it density} fluctuations by less than $O(10^{-3})$ as constrained by the CMB dipole magnitude. This is not easy to achieve for a primordial mechanism since the same mechanism responsible for an $O(10^{-1})$ (as seen by the value of $A$) modulation of the primordial density perturbation spectrum should practically leave unaffected the magnitude of these density perturbations. This however would be easier to achieve for a late time mechanism operating after the density perturbations have been created.

On smaller scales ($k\sim$ 1 Mpc$^{-1}$ there is lack of asymmetry in primordial density perturbations as constrained by the SDSS quasar sample \cite{Hirata:2009ar}. The power spectrum of these quasars is found to be isotropic with $A<0.0153$ (at $3$$\sigma$) for a dipole oriented in the direction of the CMB power \mbox {asymmetry dipole}.

Thus, the possible mechanism that leads to the observed CMB power asymmetry at the $O(10^{-1})$ level should affect the magnitude of the density perturbations at a level less than $10^{-3}$ and should also not create power asymmetry on scales $k\sim$ 1 Mpc$^{-1}$ at a level more than about $O(10^{-2})$.

Both of these constraints of this physical mechanism are consistent with a late generation of asymmetry through through large scale dark energy inhomogeneities which affect locally the growth of structures and thus the CMB maps through the ISW effect.
On the other hand, it is hard to construct primordial models of inflation that satisfy the above two constraints \cite{Dai:2013kfa,Chang:2013vla}.

It has been shown recently \cite{Rassat:2013gla} that a large part of the asymmetry is due to the ISW effect. In the context of inhomogeneous dark energy, regions of denser dark energy would be associated more pronounced ISW effect.
Thus there are two hints that point out to the possible late time origin of the CMB asymmetry:
\begin{itemize}
\item
A large part (if not all) of the asymmetry appears to be due to the ISW effect which occurs at late times and is not related to the primordial nature of the CMB perturbations \cite{Rassat:2013gla}.
\item
The power spectrum of large scale structure does not show evidence for such an dipole asymmetry on smaller scales. This also hints towards a possible late time origin of the asymmetry.
\end{itemize}

Dark energy inhomogeneities is clearly a prime candidate that could lead to the generation of late time asymmetry without affecting the spectrum of density perturbations.

%The recently observed correlation between supernovae redshifts and CMB pixel temperatures\cite{Yershov:2012sv} which is unexpectedly more pronounced at high redshifts ($z\in [0.5,1]$) may be consistent with the existence of inhomogenous dark energy which is less dense in regions where the number density of high $z$ supernovae is larger.

\subsection{Alignment of Low Multipoles in the CMB Angular Power Spectrum}

The alignment of the normals to the octopole and quadrupole planes has been verified by several studies \cite{Bielewicz:2004en,Tegmark:2003ve,Frommert:2009qw,Schwarz:2004gk,Land:2005ad,Land:2005jq,Land:2004bs}. The majority of these studies conclude that this is not a foreground contamination of the WMAP Internal Linear Combination (ILC) maps but rather is an effect degraded by the foregrounds \cite{Park:2006dv,deOliveiraCosta:2006zj}  (see however \cite{Naselsky:2003dd} for an alternative point of view). The directions of the quadrupole and the octopole are determined by rotating the coordinate system while determining the multipole coefficients $a_{lm}$. The coordinate system that maximized the sum $\vert a_{ll} \vert ^2 + \vert a_{l-l} \vert ^2$ is selected and the direction of its $z$-axis defines the direction of the $l$ multipole moment ($l=2$ for the quadrupole and $l=3$ for the octopole). This coordinate system maximizes the power of fluctuations at its $x-y$ plane (equator). The exact direction of these moments varies slightly in accordance with the method used to filter the foregrounds.

According to the recent Plack data \cite{Ade:2013nlj} and one of the filtering methods used, the  octopole plane normal is in the direction $(l,b)=(246^\circ, 66^\circ)$ , and the quadrupole plane normal is in the direction \mbox {$(l,b)=(228^\circ, 60^\circ)$}. For comparison the  CMB dipole moment is in the direction \mbox {$(l,b)=(264^\circ, 48^\circ)$ \cite{Lineweaver:1996xa}}.

The angle between the quadrupole and the octopole directions is found to in the range  $9^\circ$--$13^\circ$  depending on the foreground cleaning method used \cite{Ade:2013nlj}. The probability that this occurs by chance in a random simulation of the best-fit {\lcdm} model is found to be about $2\%$. The alignment however gets significantly degraded by subtracting any large scale specific feature (hot or cold spot) on the CMB map. Thus, it appears the there is  no specific CMB feature responsible for the alignment. Instead, the alignment appears to be an unlikely statistical fluctuation of amplitudes and phases which could be justified by the a posteriori selection of the particular statistic \cite{Bennett:2010jb}. Despite the degraded statistical significance of this anomaly by the Planck data it remains an unlikely feature in the context of {\lcdm} hinting towards statistical anisotropy. A well defined theoretical model that could naturally justify its existence would have to naturally produce a combination of large scale features in a single CMB map plane leading to the observed alignment.

\subsection{Large Scale Velocity Flows}

The kinematic Sunyaev-Zeldovich (kSZ) effect is a spectral distrortion on the CMB photons induced by the coherent velocities of the free electrons in cluster due to the peculiar velocity of the cluster as a whole. A bulk flow of clusters would produce spectral distortions at the locations of clusters whose anisotropy would be well described by a dipole. This method was first applied for the measurement of large scale bulk flows of clusters in \cite{Kashlinsky:2000xk}. It has the advantage that it is independent of other conventional methods to measure peculiar velocities which require the measurement of distances. In addition it can easily probe scales up to the horizon which is currently impossible with conventional methods. In general however it is not easy to distinguish the small signal of the kSZ dipole from the much larger signal of the intrinsic CMB dipole. Despite of these difficulties the authors of \cite{Kashlinsky:2009dw, Kashlinsky:2008ut,AtrioBarandela:2012fs} used a sample of 700 X-Ray selected clusters with redshifts up to $z\sim 0.2$ (scale of about 800 Mpc) and detected a persistent kSZ dipole which they interpreted as evidence for bulk flow of clusters with magnitude 600--1000 km/s in a direction $l \simeq 296^\circ \pm 28^\circ$, $b\simeq 39^\circ \pm 14^\circ$.

Other studies however have not confirmed this result \cite{Ade:2013opi,Hand:2012ui,Keisler:2009nw,Osborne:2010mf,Mody:2012rh} using somewhat different \mbox {methodologies and datasets}. Most of these studies \cite{Ade:2013opi,Keisler:2009nw,Mody:2012rh} have obtained a similar dipole but have claimed significantly larger errors which make the derived magnitudes of bulk flows statistically insignificant and consistent with null (and with {\lcdm} which predicts much smaller coherent peculiar velocities on these scales). The issue remains controversial \cite{Atrio-Barandela:2013ywa} and there is currently a debate as to the identification of the proper methodology for the implementation of the kSZ effect to derive the peculiar velocities of clusters.

On smaller scales (50--100 h$^{-1}$Mpc), an application of a �Minimal Variance�  weighting algorithm to a
compilation of 4481 peculiar velocity measurements \cite{Watkins:2008hf} lead to a bulk flow of 407 $\pm$ 81 km/s towards $l \simeq 287^\circ \pm 9^\circ$, $b\simeq 8^\circ \pm 6^\circ$. This result is inconsistent (too large) with respect to {\lcdm} normalized on WMAP7 at a level of $98\%$ while the direction of the flow is consistent with the result of \cite{Kashlinsky:2009dw}. A more recent analysis \cite{Turnbull:2011ty} using the peculiar velocities of a sample of 245 SnIa has obtained a bulk flow of a lower magnitude ($248 \pm 87$ km/s) towards a similar direction $l \simeq 319^\circ \pm 25^\circ$, $b\simeq 7^\circ \pm 13^\circ$. This magnitude of the bulk flow is marginally consistent with both {\lcdm} and with the previous analysis \mbox {of \cite{Watkins:2008hf}} and its direction is even closer to  other anisotropy directions discussed below (dark energy dipole, $\alpha$ dipole and CMB cold spot I \cite{Mariano:2012ia}). Consistent results with {\lcdm} are also obtained by \cite{Dai:2011xm} using SnIa peculiar velocities to measure bulk flows.

\subsection{Large Scale Alignment in the QSO Optical Polarization Data}

The optical polarization properties of 355 quasars with redshifts up to $z\sim 2.5$ and with well defined polarization angles were studied in \cite{Hutsemekers:2005iz} (see also \cite{Hutsemekers:2001xm}). The cosmic scales covered are up to about 2 Gpc. It was found that the quasar polarization vectors are not randomly distributed in the sky at a confidence level of $99.9\%$. Sources of systematics like instrumental and interstellar polarizations were carefully considered and shown not to affect the main conclusions of the analysis. In addition, the polarizations of quasars obtained by different observatories were found to agree within the uncertainties in both polarization degree and angle. The observed large scale angular correlations of quasar polarization in regions sized at least  1 Gpc  showed a coherent change of the mean polarization angle with redshift. The effect appeared stronger along an axis close to the CMB dipole even though direction uncertainties are large. If the effect is not due to an unknown systematic or due to a large statistical fluctuation, it may be  evidence for departure from the cosmological principle (large scale isotropy).

An interesting physical mechanism \cite{Hutsemekers:2005mods} that could lead to the observed coherent rotation of quasar polarizations with redshift involves a coupling between the photon and a pseudoscalar (e.g., axion) in a Lagrangian of the form:
\be
{\cal L}=\frac{1}{2}(\partial_\mu \phi) (\partial^\mu \phi) -\frac{1}{2} m^2 \phi^2 - \frac{1}{4} F_{\mu \nu}F^{\mu \nu}-j_\mu A^\mu +  \frac{1}{4}g \phi F_{\mu \nu}{\tilde F}^{\mu \nu}
\label{langpseudosc}
\ee
where ${\tilde F}^{\mu \nu}$ is the dual Maxwell tensor and $\frac{1}{4}g \phi F_{\mu \nu}{\tilde F}^{\mu \nu}$ describes the mixing pseudoscalar axions and photons. In the context of such a Lagrangian, photons with polarization parallel to a weak external magnetic field decay into pseudoscalars leading to a net polarization $p$. Thus coherent variation of the polarization $p$ over cosmological distances are predicted provided that there is a magnetic field \mbox {$B < 1$  $nG$ \cite{Payez:2008pm}.}

If there is a variation of the direction of ${\vec B}$ with distance then a rotation of the polarization angle with redshift is predicted. In an extended model where the pseudovector $\phi$ also plays the role of quintessence with induced inhomogeneities a correlation between accelerating expansion and quasar polarization would be anticipated.

Recent data showing a similar polarization alignment of radiowaves \cite{Tiwari:2012rr} on scales up to \mbox {1 Gpc} however appear to impose constraints on  the pseudoscalar coupling mechanism \cite{Payez:2012rc}. Similar strong constraints on this mechanism are imposed by the observed lack of circular polarization in the \linebreak quasar data \cite{Payez:2011sh}.

\subsection{Anisotropy in Accelerating Expansion Rate (Dark Energy Dipole)}

\begin{figure}[H]
\centering
\includegraphics[scale=1]{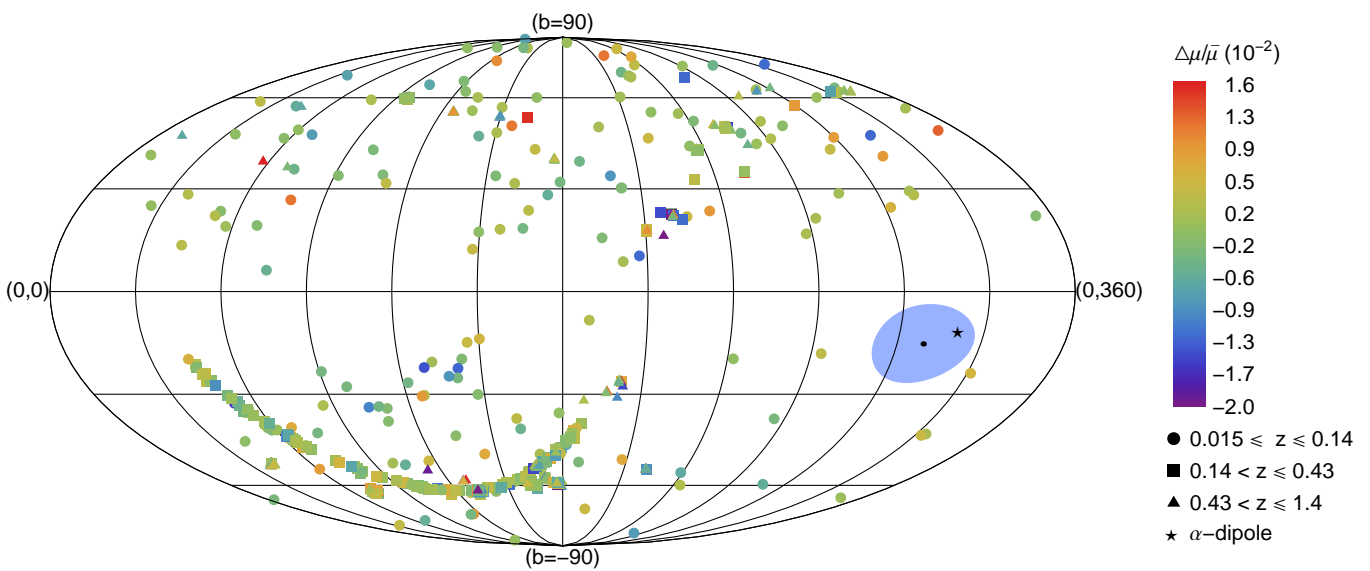}
\caption{The distribution of the Union2 SnIa datapoints in galactic coordinates along with the Dark Energy dipole direction are shown.
The datapoints are split in three different redshift bins shown with different
shapes. The direction of the $\alpha$-dipole  is also shown with a star.
The light blue blob represents the 1-$\sigma$ error on the Dark Energy dipole direction  \mbox {(from \cite{Mariano:2012wx}).}}
\label{fig:Union2data}
\end{figure}

Several recent studies \cite{Kalus:2012zu,Mariano:2012wx,Colin:2010ds,Schwarz:2007wf,Cai:2011xs,Campanelli:2010zx,Antoniou:2010gw,
Zhao:2013yaa,Cai:2013lja,Blomqvist:2008ud,Blomqvist:2010ky,Cooke:2009ws,Gupta:2010jp} have investigated the isotropy of the accelerating cosmological expansion rate at various redshift ranges using samples of SnIa as standard candles. Most studies search for a preferred direction using either a hemisphere comparison method \cite{Schwarz:2007wf,Kalus:2012zu,Antoniou:2010gw,Cai:2011xs,Gupta:2010jp} or a fit to a dipole angular distribution of the anisotropy \cite{Cooke:2009ws,Mariano:2012wx,Cai:2013lja,Campanelli:2010zx} (dark energy dipole). There are indications \cite{Mariano:2012wx} that the dipole fit may be more sensitive in identifying the expansion anisotropy signal than the hemisphere comparison method. The consensus of these studies is that there is an asymmetry axis pointing in the direction ($l\simeq 310^\circ,b=-15^\circ$) which is roughly consistent with the direction of the CMB Cold Spot I \cite{Bennett:2010jb}. This asymmetry is more prominent for low redshift SnIa ($z\leq 0.2$) perhaps due to the fact that there are significantly more and better quality SnIa data for this redshift range \cite{Mariano:2012wx}. The magnitude of the asymmetry is marginally consistent with {\lcdm} (at about $2$$\sigma$ level). The direction of this asymmetry axis appears to be consistent at $1$$\sigma$ with both the peculiar velocity flows direction and with the alpha dipole direction discussed below  (Figure \ref{fig:Union2data}).

The approximate coincidence of these four preferred directions (dark flow, dark energy dipole, $\alpha$ dipole and CMB cold spot) may hint towards a possible common physical origin for these anomalies.

\newpage
\subsection{Anisotropy in the Values of the Fine Structure Constant $\alpha$ ($\alpha$ Dipole)}

The absorption spectra of distant quasars contain two pieces of information: the redshift of the absorber and the value of the fine structure constant $\alpha$. This information extracted by the Many Multiplet Method has been used to perform an anisotropy analysis of the fine structure constant $\alpha$ \cite{King:2012id,King:2012id2}. The analysis is based on a large
sample of quasar absorption-line spectra (295 spectra) obtained using UVES (the Ultraviolet and Visual
Echelle Spectrograph) on the VLT (Very Large Telescope) in Chile and also previous observations at the Keck Observatory in Hawaii.

An apparent variation of $\alpha$ across
the sky was found. It was shown to be well fit  by an angular dipole model $\daa=A \cos\theta + B$ where $\theta$ is the angle with respect to a preferred axis and $A,B$ are the dipole magnitude and an isotropic monopole term. The dipole axis was found to point in the direction
$(l,b)= (331^\circ, -14^\circ)$ and the dipole amplitude $A$ was found to be $A = (0.97\pm 0.21)\times 10^{-5}$. The statistical significance over an isotropic model is at the $4.1\sigma$ level. The analysis of \cite{King:2012id,King:2012id2} has received criticism \cite{carroll,carroll1,carroll2} based mainly on the fact that its quasar sample combines two datasets (Keck and VLT) with different systematic errors  which have a small overlapping subset and cover opposite hemispheres on the sky. The axis connecting these two hemispheres has similar direction with the direction of the obtained dipole. The response of the authors of ~\cite{King:2012id,King:2012id2} was based on the fact that in the equatorial region of the dipole,
where both the Keck and VLT samples contribute a number of absorbers, there is no evidence for inconsistency between Keck and VLT.

The angular separation between the Dark Energy Dipole direction and the $\alpha$ dipole is found to be only $11.3^\circ \pm 11.8^\circ$. Reference \cite{Mariano:2012wx} used Monte Carlo simulations to find the probability of obtaining the observed dipole magnitudes with the observed alignment, in the context of an isotropic cosmological model with no correlation between dark energy and fine structure constant $\alpha$. It was found that this probability is less than one part in $10^6$.

 A simple physical model (extended topological quintessence \cite{BuenoSanchez:2011wr,Mariano:2012wx}) naturally predicts a spherical inhomogeneous distribution for both dark energy density and fine structure constant values. The model is based on the existence of a recently formed giant global monopole with Hubble scale core which also couples non-minimally to electromagnetism. Aligned dipole anisotropies would naturally emerge for an off-centre observer for both the fine structure constant and for dark energy density. This model smoothly reduces to {\lcdm} for proper limits of its parameters. Two predictions of this model are (a) a correlation between the existence of strong cosmic electromagnetic fields and the value of $\alpha$ and (b) the existence of a dark flow on Hubble scales due to the repulsive gravity of the global defect core (``Great Repulser'') aligned with the dark energy and $\alpha$ dipoles. The direction of the dark flow is predicted to be towards the spatial region of lower accelerating expansion. Existing data about the dark flow \cite{Watkins:2008hf,Kashlinsky:2008ut} are consistent with this prediction. In the next section we present a detailed analysis of this model.

 \subsection{Large Quasar Groups}

According to SDSS DR7 for galaxies with $0.22 < z < 0.50$ \cite{Marinoni:2012ba}, isotropy appears to be established on scales larger than 210 Mpc as anticipated from the overall isotropy of the CMB. However, on larger redshifts, isotropy appears to be challenged by recent quasar data.

The Large Quasar Groups (LQG) are large structures of quasars appearing at redshifts $z~1$ with scales 70--350 Mpc.  A new elongated LQG was discovered recently \cite{Clowes:2012pn} in the DR7QSO catalogue of the Sloan Digital Sky Survey \cite{Schneider:2010hm}. It has long dimension 1240 Mpc and characteristic size $(Volume)^{1/3}$ about 500 Mpc  at a redshift $z= 1.27$ (Huge-LQG). It is located in the direction $l=225^\circ$, $b=63^\circ$, relatively close (distance of boundaries  140 Mpc) to another fairly large LQG and consists of 73 member quasars. The principal axes of the LQG  have lengths of  1240, 640, and  370 Mpc extending to the Gpc-scale. According to   \cite{Yadav:2010cc}, the upper limit to the scale of homogeneity for the concordance cosmology is  370 Mpc. Their long dimensions from the inertia tensor of  630 and  780 Mpc are clearly much larger than the anticipated upper limit of homogeneity. It is the currenly known largest structure in the universe and maybe at odds with the cosmological principle \cite{Yadav:2010cc}.  We stress however that other more recent studies have indicated \cite{Nadathur:2013mva,Pilipenko:2013yna} that such structures may not be inconsistent with a homogeneous Poisson distribution of quasars. Indeed, such structures seem to be discovered  \cite{Nadathur:2013mva,Pilipenko:2013yna} by the algorithm \mbox {of \cite{Clowes:2012pn}} in  explicitly homogeneous simulations of a Poisson point process with the same density as the \mbox {quasar catalogue.}

\section{Anisotropic/Inhomogeneous Dark Energy Models}

In view of the successes of the standard {\lcdm} model it is anticipated that a viable theoretical model able to explain the above described anomalies should have the following properties:
\begin{itemize}
\item
Deviate from {\lcdm} mainly on large cosmological scales.
\item
Reduce to {\lcdm} for certain values of its parameters.
\end{itemize}

In addition, it appears reasonable and in fact hinted by some of the above described data that the physical origin of the deviation from isotropy resides at late cosmological times when these large scales are again causally connected (consider, e.g., the possible ISW origin of the CMB power \mbox {asymmetry \cite{Rassat:2013gla}}). Motivated by these arguments and by the discussion presented in the Introduction we focus on models involving anisotropic and/or inhomogeneous dark energy. In contrast to matter, dark energy can be anisotropic (due to its pressure) without being inhomogeneous and also the possible inhomogeneities of dark energy naturally appear on Hubble scales where the origin of the anomalies seems to reside. In addition, Gpc  inhomogeneities (voids) of matter constitute models that do not smoothly reduce to {\lcdm} for any value of their parameters. This class of models has been shown to be severely constrained by several cosmological observations and especially the kSZ effect which is predicted to be significantly larger in these models than observations indicate \cite{Caldwell:2007yu}. Thus in what follows we focus on models of anisotropic/inhomogeneosu dark energy.

\subsection{Anisotropic Dark Energy Models}

A generic generalization of the Friedmann-Lema\^itre-Robertson-Walker (FLRW) metric  that \mbox {introduces} anisotropy but not inhomogeneity is the Bianchi I metric which is of the form:
\be \label{b1metric}
ds^2 = -dt^2 +a^2(t)dx^2+b^2(t)dy^2 + c^2(t)dz^2
\ee

Such a metric is consistent with an energy momentum tensor of a fluid with homogeneous but anisotropic  pressure of the form:
\be
T^\mu_{\nu} = diag(-1,w_x,w_y,w_z)\rho
\label{enmom1a}
\ee
where $(w_x,w_y,w_z)$ are the equation of state componenets corresponding to each component of pressure. The deviation of these components may be parametrised by two parameters \cite{Koivisto:2007bp} $\gamma$ and $\delta$. Using these parameters, the anisotropic dark energy energy momentum tensor Equation  (\ref{enmom1a}) may be expressed as:
\be
T^\mu_{\phantom{\mu}\nu} = diag(-1,w,w+3\delta,w + 3\gamma)\rho
\ee

The deviations from isotropy may be quantified by defining the average Hubble expansion rate $H$ and the Hubble normalized shear $R$ and $S$ as:
\be
H \equiv \frac{1}{3}\left(\frac{\dot{a}}{a}+\frac{\dot{b}}{b} + \frac{\dot{c}}{c}\right)
\ee
and
\be
R  \equiv \frac{1}{H}\left(\frac{\dot{a}}{a}-\frac{\dot{b}}{b}\right), \quad
S  \equiv \frac{1}{H}\left(\frac{\dot{a}}{a}-\frac{\dot{c}}{c}\right)
\ee

The generalized Friedmann equation is then obtained as:
\be
H^2 = \frac{8\pi G}{3}\frac{\rho + \rho_m}{1-\frac{1}{9}\left(R^2+ S^2 - RS\right)}
\ee
where $\rho_m$ is the matter density which in general is allowed to interact with dark energy.
Conservation of the energy momentum tensor of anisotropic dark energy implies:
\be
\dot{\rho}   + 3\left[\left(1+w\right)H + \delta \frac{\dot{b}}{b} +
\gamma\frac{\dot{c}}{c} \right]\rho = -Q H \rho_m
\ee
where $\rho_m$ is the matter density which obeys:
\be
\dot{\rho}_m + 3H\rho_m = Q H \rho
\ee
and a nonzero value of the parameter $Q$ expresses the interaction between dark energy and matter. Such an anisotropic dark energe could emerge for example in the presence of a vector field $A_\mu$ whose dynamics is described by the action \cite{Koivisto:2007bp}:
\be\label{vector}
S = \int d^4 x \sqrt{-g} \left[\frac{R}{16\pi G} -\frac{1}{4}F_{\mu\nu}F^{\mu\nu} - V \right]
\ee
where  $F_{\mu\nu} = \partial_\mu A_\nu - \partial_\nu A_\mu$ and the potential $V$ is a function of $A^2 = A_\mu A^\mu$ \cite{Koivisto:2007bp}.

The main observational effect of anisotropic dark energy is its contribution to the CMB anissotropies at the level of low multipole moments. This contribution is introduced through the last scattering redshift which becomes anisotropic in the context of the metric Equation (\ref{b1metric}).
For a photon observed from the a direction $\hat{{\bf p}}$ the last scattering redshift $z$ may be obtained using the lightlike geodesic equatios. It is of the form:
\be
\label{z_ellips}
1+z(\hat{{\bf p}}) =
%\left( \sum_i \frac{\hat{p}^2_i}{a_i^2}\right)^\frac{1}{2}=
\frac{1}{a}\sqrt{1 + \hat{p}_y^2e_y^2 + \hat{p}_z^2e_z^2}
\ee
where:
\be \label{ellips}
e_y^2 = \left(\frac{a}{b}\right)^2-1, \quad
e_z^2 = \left(\frac{a}{c}\right)^2-1
\ee
are evaluated at the time of last scattering and the scale factors are all normalized to unity
today. Thus, the predicted CMB temperature is of the form:
\be
T(\hat{{\bf p}}) = T_*/(1+z(\hat{{\bf p}}))
\ee
where $T_*$ is the temperature at decoupling.
Using the spatial average $4\pi\bar{T} = \int d\Omega_{\hat{{\bf p}}}T(\hat{{\bf p}})$,
the temperature anisotropies is expressed as:
\be
\frac{\delta T(\hat{{\bf p}})}{\hat{T}} = 1- \frac{T(\hat{{\bf p}})}{\bar{T}}
\ee
This anisotropy may be expanded in spherical harmonics to evaluate the contribution to all CMB multipole moments and thus impose constraints on the parameters $\gamma$ and $\delta$ by demanding that these multipoles do not exceed the observed values. Significant fine tuning is required on the anisotropy parameters to secure consistency with the observed low quadrupole value \cite{Koivisto:2007bp,Mota:2007sz,Koivisto:2008ig}.

Less stringent constraints on the anisotropy parameters may be obtained by considering the isotropy of the accelerating cosmic cosmic expansion rate as determined using the luminosity distances of SnIa. The predicted values of luminosity distances in the context of the metric Equation (\ref{b1metric}) is obtained \cite{Koivisto:2007bp} at the redshift $z$ in the direction $\hat{p}$ as:
\be \label{lumi}
d_L(z,{\bf \hat{p}}) = (1+z)
\int_{t_0}^{t(z)}\frac{dt}{\sqrt{\hat{p}_x^2a^2+\hat{p}_y^2b^2+\hat{p}_z^2c^2}}
\ee
where $z$ is connected with the scale factors via
Equation (\ref{z_ellips}). Using the maximum likelihood method and the observed luminosity distances it is possible to obtain constraints on the anisotropy parameters $\gamma$ and $\delta$ and show that their absolute values are constrained to be less than about $0.1$ at the $3$$\sigma$ level \cite{Koivisto:2007bp}.

An alternative parametrization of the dark energy anisotropy involves the consideration the anisotropic stress $\sigma$. This is obtained by expressing the dark energy momentum tensor as \cite{Mota:2007sz}:
\be
T_{\mu\nu}= \rho u_\mu u_\nu + p (g_{\mu \nu} + u_\mu u_\nu) + \Sigma_{\mu \nu}
\ee
where where $u_\mu$ is the four-velocity of the fluid, and $\Sigma_{\mu \nu}$ is the anisotropic stress tensor. The anisotropic stress $\sigma$ is defined as:
\be
(\rho + p) \sigma = -({\hat k}_i {\hat k}_j - \frac{1}{3} \delta_{ij}) \Sigma_{ij}
\ee
The anisotropic stress $\sigma$ quantifies the deviation of pressure from isotropy. As discussed in  \cite{Koivisto:2007bp}, only weak constraints can be imposed on $\sigma$ using recent data.

The introduction of anisotropic dark energy as described above provides the potential for breaking of statistical isotropy and resolution of some of the large scale anomalies discussed in the previous section. In particular,  the anisotropic pressure can induce peculiar velocity flows, anisotropy in the SnIa data and significant CMB dipole and quadrupole \cite{Koivisto:2007bp}. The most severe constraints on the level of anisotropy originate from the CMB observed isotropy but these can be softened by considering a time dependent anisotropy which is more significant at late times and consistent with the weaker constraints originating from the SnIa data.

\subsection{Inhomogeneous Dark Energy Models}

In the context of inhomogeneous cosmologies, the simplest way to introduce a preferred direction in the universe is to consider a Hubble scale spherical dark matter void or dark energy inhomogeneity and an off center observer. In the case of dark matter void, an observer close to the center of the void will detect a larger Hubble parameter for nearby points in space and lower redshift. Thus she may incorrectly interprete it as an accelerating expansion rate since the Hubble expansion will appear higher at lower redshifts. Thus a spatial variation of expansion rate may be misinterpreted as a temporal variation. In the case of a spherical dark energy overdensity the variation of the expansion rate is both spatial and temporal and there is a true accelerating expansion rate amplified by the spatial variation of the dark energy density.

\begin{figure}[H]
\centering
\includegraphics[trim = 0mm 240mm 115mm 10mm, clip=true, width=0.9\textwidth]{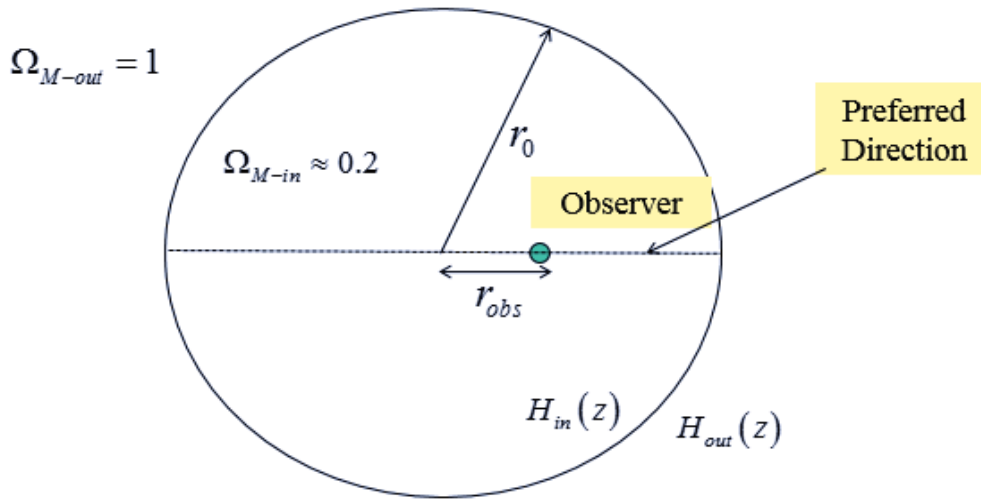}
\caption{An off-center observer in a spherical inhomogeneity sees a preferred direction of higher accelerating expansion rate.\vspace{-24pt}} \label{void-displ}
\end{figure}

The displacement direction will be naturally preferred and the observer will experience anisotropic accelerated Hubble expansion and Hubble scale velocity flows (Figure \ref{void-displ}). In fact the center of the dark matter void or dark energy overdensity will behave as a great repulser inducing velocity flows away \mbox {from it.}

\vspace{-8pt}
As a simple toy model we may consider an inhomogeneous fluid of matter and/or dark energy described by the energy-momentum tensor:
\be
T_\nu^\mu = \diag(-\rho(r)-\rho_M(r,t),p_r(r),p_t(r),p_t(r))
\label{enmom1}
\ee

The general spherically symmetric inhomogeneous cosmological spacetime is described by the metric~\cite{Lemaitre:1933gd,Tolman:1934za,Bondi:1947}:
\begin{equation}
ds^2 = - dt^2 + A^{2}(r,t)dr^2 + B^{2}(r,t) \left( d\theta^2 +
\sin^2 \theta d\varphi^2 \right)
\label{ltbmet}
\end{equation}

In the special case of a diagonal energy momentum tensor as the one of Equation  (\ref{enmom1}), the above metric takes the simplified form \cite{Lemaitre:1933gd,Tolman:1934za,Bondi:1947}:
\begin{equation}\label{ltbmet2}
ds^2 = - dt^2 + \frac{(B'(r,t))^2}{1-k(r)}dr^2 + B^{2}(r,t) \left(
d\theta^2 + \sin^2 \theta d\varphi^2 \right)\,
\end{equation}
where $B(r,t)$ is the spacetime dependent scale factor and $k(r)$ plays the role of inhomogeneous spherically symmetric spatial curvature. In the limit of a homogeneous spacetime, the LTB metric Equation  (\ref{ltbmet2}) reduces to the FLRW metric by setting $k(r)$ to a constant and $B(r,t)= r a(t)$. Using the metric Equation  (\ref{ltbmet2}) and the energy momentum tensor Equation  (\ref{enmom1}) in the Einstein equations we obtain the generalized \mbox {Friedman equation}:
\begin{equation}\label{freq3}
H^2(r,t) = H_0^2(r) \left[ \Omega_M(r) \left(\frac{B_0}{B} \right)^3
+ \Omega_X(r) + \Omega_c(r) \left(\frac{B_0}{B} \right)^2
\right]
\end{equation}
where:
\begin{equation}\label{hupple}
H(r,t) \equiv \frac{\dot{B}(r,t)}{B(r,t)}
\end{equation}
with $B_0(r)\equiv B(r,t_0)$ ($t_0$ is the present time), $H_0(r) \equiv H(r,t_0)$,
$\Omega_X (r) \equiv -{8 \pi G p_r(r)}/{3 H_0^2(r)}$ (which is positive for negative radial pressure) and $\Omega_c(r) \equiv 1- \Omega_X (r)-\Omega_M(r)$. The profile of the inhomogeneous dark energy pressure, $\Omega_X(r)$, is in principle arbitrary, and is physically determined by the physical mechanism that induced the inhomogeneous overdensity.

Interesting density profiles of the inhomogeneities (dark matter and dark energy) which parametrize the inside and outside densities ($\Omega_{ \text{in}}$ and $\Omega_{\text{out}}$) as well as the interface thickness ($\Delta r$) are of the form:
%\begin{widetext}
\ba \label{omegam}
   \Omega_M(r)&=&\Omega_{M, \text{out}}+(\Omega_{M, \text{in}}-\Omega_{M, \text{out}}) \, \frac{1-\tanh [(r-r_0) / 2\Delta r ]}{1+\tanh (r_0/ 2\Delta r )} \\
 \label{omegax}
   \Omega_X(r)&=&\Omega_{X, \text{out}}+(\Omega_{X, \text{in}}-\Omega_{X, \text{out}}) \, \frac{1-\tanh [(r-r_0) / 2\Delta r ]}{1+\tanh (r_0/ 2\Delta r )} \,
\ea
%\end{widetext}

It is straightforward to solve Equation  (\ref{freq3}) by first determining $H_0(r)$ in units of Gyr$^{-1}$, assuming an age of the universe $t_0=13.7$\, Gyr (homogeneous Big Bang). We thus use:
\begin{equation} \label{age}
   H_0(r)=\frac{1}{t_0}\int\limits^1_0 \frac{dx}{\sqrt{\Omega_M(r)x^{-1}+\Omega_X(r)x^2+\Omega_c(r)}}
\end{equation}
which can be easily derived from
Equation~(\ref{freq3}). With
Equation (\ref{age}) equation and the initial condition $B(r,t_0)=r$ we solve
Equation ~(\ref{freq3}) and obtain numerically $B(r,t)$ and its derivatives.

For an observer in the center of the spherical inhomogeneity, light travels radially inwards and therefore for these lightlike geodesics $d\theta = d\phi =0$ along the geodesic. The remaining two light-like geodesics for radially {\it incoming} light rays are of the form~\cite{Enqvist:2006cg}:
\begin{equation}\label{dtdz}
\frac{dt}{dz}=-\frac{B'(r,t)}{(1+z)\dot{B}'(r,t)}
\end{equation}
\begin{equation}\label{drdz}
\frac{dr}{dz}= \frac{c\sqrt{1-k(r)}}{(1+z)\dot{B}'(r,t)}\
\end{equation}
where $c\simeq 0.3$ is the velocity on light in units of Gpc/Gyr used for consistency in order to \linebreak obtain the luminosity distance in Gpc. Using these lightlike geodesics with initial conditions \linebreak $t(z=0)=t_0$ and $r(z=0)=0$ we find $t(z)$ and
$r(z)$. The angular diameter and luminosity distances are then obtained as:
\begin{eqnarray}
d_{A}(z,r_0,\Omega_{X, \text{in}}) &=& B(r(z),t(z)) \,\\
d_{L}(z,r_0,\Omega_{X, \text{in}}) &=& (1+z)^{2} \, d_{A}(z) \label{lumidi} \,
\end{eqnarray}

\begin{figure}[H]
\centering
\includegraphics[trim = 0mm 190mm 100mm 10mm, clip=true, width=0.9\textwidth]{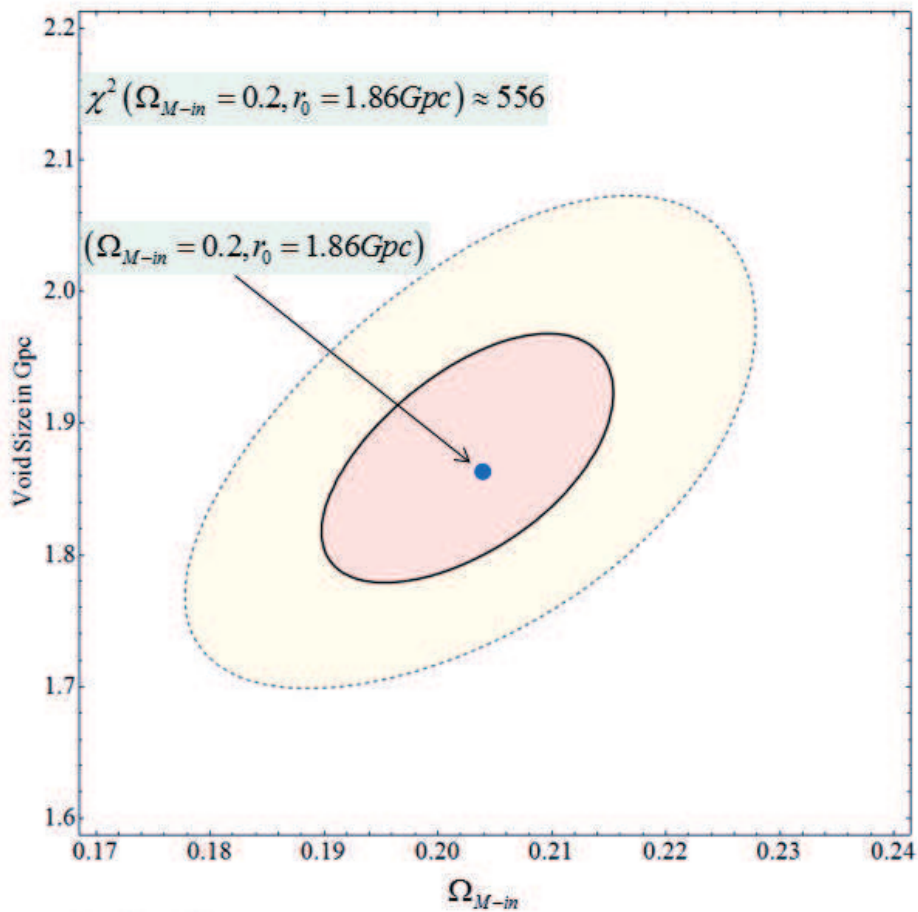}
%\rotatebox{0}{\hspace{0cm}\resizebox{1.7 \textwidth}{!}{\includegraphics{void-params.eps}}}
\hspace{0pt}
\caption{The $1$$\sigma$ and $2$$\sigma$ contours for the parameters $\Omega_{M,\text{in}}$ (matter underdensity inside the void) and $r_0$ (size of the void) (from   \cite{Grande:2011hm}). \vspace{-18pt} \label{voidparams} }
\end{figure}

Using the above form of the luminosity distance, it is straightforward to fit the predicted distance moduli to the observed ones from the Union2 dataset and derive the best-fit values of parameters for the on-center observer. We first consider a dark matter inhomogeneity (void) with no dark energy. Fixing the interface thickness to $\Delta r = 0.35$ Mpc, $\Omega_{X,\text{in}}=\Omega_{X,\text{out}}=0$ (no dark energy), $\Omega_{M,\text{out}}=1$, it is straightforward to obtain \cite{Grande:2011hm} the $1$$\sigma$ and $2$$\sigma$ contours for the parameters $\Omega_{M,\text{in}}$ (matter underdensity inside the void) and $r_0$ (size of the void) shown in Figure \ref{voidparams}.

\vspace{-24pt}

This class of Gpc void models had until recently attracted significant attention. They have three \linebreak main advantages:
\begin{itemize}
\item
They are consistent with the observed redshift dependence of the Hubble expansion rate as obtained using standard candles SnIa \cite{Enqvist:2006cg,Alnes:2006uk} without the requirement of dark energy.
\item
They can naturally provide a preferred axis by utilizing a slight displacement of the observer from the center of the spherical void.
\item
The coincidence problem is not present since a void can naturally develop at late times along with other structures.
\end{itemize}

Despite of the above successes, it has been evident that this class of models also suffers from significant problems. As discussed in the Introduction, these problems include the following:
\begin{itemize}
\item
In the context of standard cosmology there is no simple mechanism to create Gpc matter voids without conflicting other cosmological data (e.g., CMB fluctuations).
\item
If the observer location is not fine tuned close to the center of the spherical void, then a CMB dipole is predicted which is larger than the observed \cite{Alnes:2006pf,Grande:2011hm}.
\item
There is no simple matter density profile that will provide an equal or better quality of fit compared to {\lcdm} \cite{Grande:2011hm}.
\item
This class of models generically predicts large peculiar velocities of free electrons and clusters with respect to the CMB. These velocities could have been detected via inverse Compton scattering of CMB photons onto moving free electrons (kinematic Synyaev-Zeldovich effect). The existing bounds on such velocities effectively rules out this class of models \cite{Caldwell:2007yu,GarciaBellido:2008gd}.
\end{itemize}

In view of the above problems of matter void models (see also \cite{GarciaBellido:2008nz}), it is of some interest to consider models where the Gpc void is replaced by a dark energy spherical overdensity with similar size. As discussed in the Introduction, the motivation for this class of inhomogeneous dark energy models may be summarized as follows:
\begin{itemize}
\item
The standard cosmological model {\lcdm} assumes the existence of homogeneous matter on large scales, the validity of general relativity and the existence of homogeneous dark energy with constant in time energy density (cosmological constant). The generalization of most of these assumptions has been extensively studied in the literature. The main motivation for such generalizations originates at the coincidence problem expressed as a ``Why now?'' problem: ``Why was the energy scale of the cosmological constant tuned to such low values so that it starts dominating at the present time?''. For example general relativity has been generalized to various models of modified gravity, dark energy was allowed to evolve in time as a scalar field (quintessence), matter was allowed to be inhomogeneous on Gpc scales (void models). The least studied generalization is the one based on allowing dark energy to become inhomogeneous on Gpc scales. Such a generalization could in principle address the coincidence problem if the later is expressed as a ``Why here?'' problem: ``Why is the dark energy density in our horizon such that the universe has recently started its accelerating expansion?''. In the same manner that the answer to the ``Why now?'' question could be time dependence, the answer to the ``Why here?'' question could be spatial dependence of dark energy density. Therefore, the consideration of inhomogeneous dark energy constitutes a generic generalization of the standard {\lcdm} models which reduces back to {\lcdm} for a inhomogeneity scale that exceeds the current horizon scale. It is therefore interesting to use cosmological observations to impose constraints on the two basic parameters of this class of models: the scale of the inhomogeneity and the magnitude of the dark energy density inhomogeneity.
\item
As in the case of the dark matter void, an off-center observer in a spherically symmetric dark energy inhomogeneity will naturally experience an anisotropy in the accelerating expansion rate. In the case of inhomogeneous dark energy the scale of the inhomogeneity is allowed to be as large as the horizon scale ({\lcdm} limit). For large enough inhomogeneity scale, the observer is allowed to be displaced significantly from the center without experiencing a large CMB dipole. Thus, inhomogeneous dark energy models are less subject to fine tuning with respect to the location of the observer \cite{Grande:2011hm}.
\item
Due to negative pressure and speed of sound close to unity, dark energy inhomogeneities are not easy to sustain. However, there is a well defined physical mechanism (topological \mbox {quintessence \cite{BuenoSanchez:2011wr})} that can lead to sustainable large scale dark energy inhomogeneities, supported by topological considerations. This mechanism can be viewed as a generalization of topological inflation \cite{Vilenkin:1994pv} in which early time accelerating expansion (inflation) takes place in the core of a topological defect due to the topologically trapped vacuum energy.
\end{itemize}

In this brief review I focus on spherically symmetric dark energy inhomogeneities corresponding to global monopoles in the context of topological quintessence. It is straightforward to generalize this assumption by considering other models that are both inhomogeneous and anisotropic. For example the Szekeres� quasi-spherical models \cite{sz1,sz2},  are effectively non-linear FLRW perturbations \cite{Goode:1982pg}, with  metric of the form:
\be
ds^2 = -dt^2 + e^{2A} dx^2 + e^{2B} (dy^2 + dz^2)
\ee
where $A = A(t, x, y, z)$, $B = B(t, x, y, z)$. Such models models can be matched to a dust FLRW model across a matter-comoving spherical junction surface \cite{Bonnor:1976zz} and may represent different types of topological defects in the context of topological quintessence (strings or walls).

\subsection{Topological Quintessence}

Topological quintessence \cite{BuenoSanchez:2011wr} assumes the recent formation of a global topological defect (e.g., a global monopole) with Hubble scale core and with core vacuum energy dansity of the same order as the current density of matter. The dynamics of such a global monopole is determined by the action:
\begin{equation}\label{gmaction}
  S=\int d^4 x \sqrt{-g} \left[\frac{m_{Pl}^{~2}}{16\pi}{\cal R}
     -\frac12(\partial_{\mu}\Phi^a)^2-V(\Phi)  + {\cal L}_m \right]\,
\end{equation}
where $ {\cal L}_m$ is the Lagrangian density of matter fields,  $\Phi^a ~(a=1,2,3)$ is an $O(3)$ symmetric scalar \linebreak field and:
\begin{equation}\label{gmptl}
V(\Phi)= {1\over 4}\lambda(\Phi^2-\eta^2)^2, ~~
\Phi\equiv\sqrt{\Phi^a\Phi^a}\,
\end{equation}

The vacuum energy density in the monopole core and the size of the core are determined by the two parameters of the model
$\eta$ (the vacuum expectation value) and $\lambda$ (the coupling constant).  The global monopole field configuration is described by the hedgehog ansatz:
\begin{equation}\label{glmonanz}
\Phi^a=\Phi(r,t)\hat r^a\equiv
\Phi(r,t)(\sin\theta\cos\varphi,\sin\theta\sin\varphi,\cos\theta)
\end{equation}
with boundary conditions:
\begin{equation}
\Phi(0,t)=0,  \Phi(\infty,t) =\eta\,
\end{equation}
where a time dependence is allowed, having in mind a cosmological setup of an expanding background.

The general spherically symmetric spacetime around a global monopole may be described by a metric of the form Equation (\ref{ltbmet}).
In order to derive the cosmological dynamical equations we need the total energy momentum tensor $T_{\mu \nu}$. We assume a cosmological setup at recent cosmological times and therefore $T_{\mu \nu}$ is dominated by the global monopole vacuum energy and matter, {\em i.e}.:
\be
	T_{\mu\nu} = T_{\mu\nu}^{(mon)} + T_{\mu\nu}^{(mat)}\,
\label{tmntot}
\ee

The energy momentum tensor of the monopole is given by:
\begin{equation}\label{ein}
T_{\mu\nu}^{(mon)}=\partial_{\mu}\Phi^a\partial_{\nu}\Phi^a
-g_{\mu\nu}\Bigl[\frac12(\partial_{\sigma}\Phi^a)^2+V(\Phi)\Bigr]\,
\end{equation}

\begin{figure}[H]
\centering
\includegraphics[trim = 0mm 220mm 110mm 10mm, clip=true, width=0.82\textwidth]{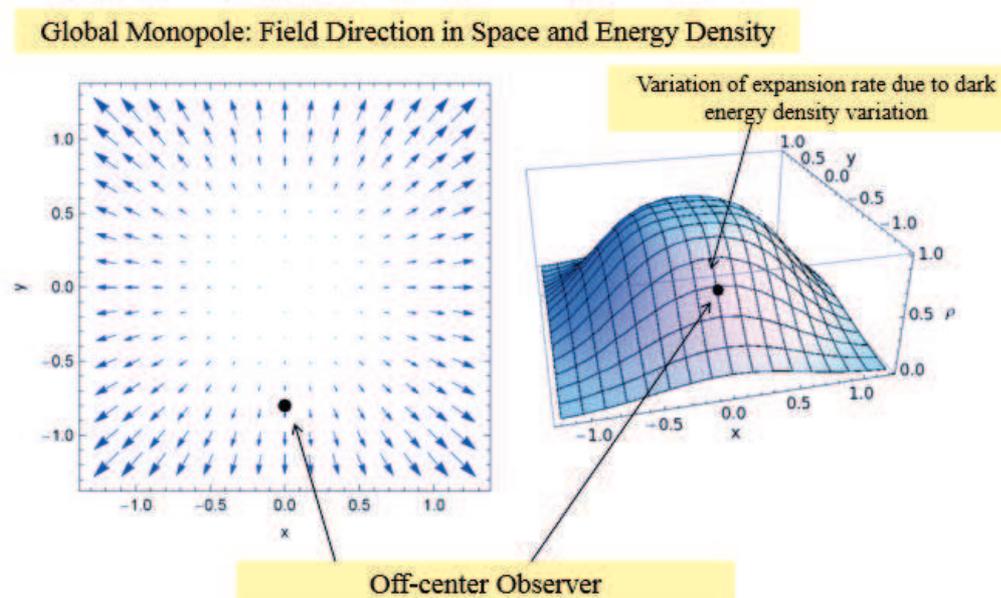}
%\rotatebox{0}{\hspace{0cm}\resizebox{1.7 \textwidth}{!}{\includegraphics{glmon.eps}}}
\hspace{0pt}
\caption{An off-center observer in the core of a global monopole will naturally experience an anisotropy in the expansion rate due to the higher vacuum energy density in the direction of the monopole center. \vspace{-12pt} \label{glmon}}
\end{figure}

The negative pressure in the monopole core is expected to lead to accelerating expansion when the core size is of the order of the Hubble scale \cite{Cho:1997rb}.
The monopole energy density is of the form:
\be
\rho^{mon}=T_{00}^{mon}=\left[ {\dot \Phi ^2 \over 2} + {\Phi'^2 \over 2A^2} +{\Phi^2 \over B^2 r^2} + {\lambda \over 4} \left(\Phi^2 - \eta^2 \right)^2 \right]\,
\label{rhomon}
\ee
and has a local maximum at the monopole core where vacuum energy dominates.
The energy-momentum tensor of matter is that of a perfect fluid with zero pressure ($p=0$):
\be
T_{\mu\nu}^{(mat)} = \rho^{mat} u_{\mu} u_{\nu}\,
\label{tmnmat}
\ee
where $\rho^{mat}=\rho^{mat}(r,t)$ is the matter density and $u^{\mu}$ is the velocity 4-vector of the fluid:
\be
u^{\mu} = ({1 \over \sqrt{1-v^2}}, {v \over A \sqrt{1-v^2}}, 0, 0)\,
\ee
with $v=v(r,t)$ the radial matter fluid velocity. An off-center observer in the global monopole will experience anisotropic accelerating cosmological expansion which is maximized in the direction of the monopole center where the vacuum dark energy density is maximum (Figure \ref{glmon}). This is consistent with the mild evidence ($2$$\sigma$) for the existence of a dark energy dipole discussed in Section 2.

\vspace{-30pt}
The question that arises is ``Can this mechanism also predict the existence of a fine structure constant dipole aligned with the dark energy dipole as seems to be favored by some observations?'' The answer to this question is positive. An $\alpha$ dipole aligned with the dark energy dipole can be obtained by considering a dilatonic coupling of the monopole scalar field to electromagnetism. Consider the generalization of the Lagrangian Equation (\ref{gmaction}) of the form:
\be
 S =\int \left[ \frac{1}{2}M_p^2 R -\frac12(\partial_{\mu}\Phi^a)^2
 -V(\Phi)  + \frac{1}{4}B(\Phi) F_{\mu\nu}^2 + {\cal L}_m
 \right]\sqrt{-g} d^4 x
\ee
where $M_p^{-2}=8\pi G$ is the reduced Planck mass. This action is inspired from the \mbox {Bekenstein-Sandvik-Barrow-Magueijo}
class of models \cite{Bekenstein:1982eu,Sandvik:2001rv,Avgoustidis:2013bqa} where the dynamics are attributed to a homogeneous scalar field. In our case (extended topological quintessence) the fine structure constant $\alpha$ has obtained dynamics through the inhomogeneous global monopole scalar field. It may be shown that:
\be
\alpha(\Phi)=\frac{e_0^2}{4\pi B(\Phi)^2}
\label{alphabphi}
\ee
where $e_0$ is the bare charge that remains constant throughout the cosmological evolution. By considering a dilatonic coupling $B(\Phi)$ of the form
Consider now a non-minimal coupling of the form:
\be
B(\Phi)=1-\xi \frac{\Phi^2}{\eta^2}
\label{nmcoupl}
\ee
where $\xi$ is constant, we find for small values of $\Phi/\eta$ a spatial variation of $\alpha$ in accordance to:
\be
\daa \simeq 2\xi \frac{(\Phi^2-\Phi_0^2)}{\eta^2}
\label{daaphi1}
\ee
where $\Phi_0$ is the field magnitude at the location of the observer. Clearly, $\alpha$ is predicted to vary most in the direction where the field magnitude varies more rapidly. This is the direction towards the center of the monopole which is identical to the direction of the dark energy dipole. Thus, this model naturally predicts the observed alignment between the $\alpha$ dipole and the dark energy dipole.

The observationally required values of the parameters of topological quintessence may be obtained by implementing the requirement that the core size is comparable  to the Hubble scale and the vacuum energy density at the monopole core is comparable to the matter density at present. The core scale (where $\Phi/\eta \lsim 1$) of the static global monopole is:
\be \delta \simeq \lambda^{-1/2} \eta^{-1}\, \label{coresc} \ee while the vacuum energy density in this core region is:
\be \rho^{core} \simeq \frac{\lambda \eta^4}{4}\, \label{rhomoncore} \ee

Deep inside the monopole core the expansion is approximately isotropic ($A(r,t)\simeq B(r,t) \equiv a_{in}(t)$) and the dynamics of the scale factors are approximated by:
\be
3 H_{in}(t)^2 = {8\pi \over m_{Pl}^2} \l( {\lambda \eta^4 \over 4} + \rho^{mat}_{0,in} \l( {a_{0,in} \over a_{in}}\r)^3 \r)\,
\label{lcdmmon}
\ee
where $a_{0,in}\equiv a_{in}(t_0)$ is the scale factor at the present time $t_0$, $\rho^{mat}_{0,in}$ is the present density of matter in the monopole core and:
\be
H_{in}(t)\equiv \l({{\dot a_{in}} \over a_{in}}\r)\,
\label{hofft}
\ee

Similarly, far away from the monopole core ($r\gg \delta$), spacetime is also approximately homogeneous and we may set $A(r,t)\simeq B(r,t) \equiv a_{out}(t)$. In this limit the dynamics of the scale factors are \mbox {obtained from:}
\be
3 H_{out}(t)^2 = {8\pi \over m_{Pl}^2}  \rho^{mat}_{0,out} \l( {a_{0,out} \over a_{out}}\r)^3\,
\label{matmon}
\ee
corresponding to a flat matter dominated universe.
By imposing the physical requirements stated \linebreak above namely:
\ba
\delta & \simeq & H_0^{-1} \equiv H_{in}(t_0)^{-1} \label{width} \\
\rho^{core} & \simeq & \rho^{mat}(t_0)   \label{rcore}
\ea
where $\rho^{core}$ is the vacuum energy density in the monopole core, we obtain order of magnitude estimates for the two parameters in the global monopole action:
\ba
\eta &\sim& O(1) \times M_{pl} \\
\lambda &\sim& 10^{-120} {\bar \eta}
\ea
where $\bar \eta \equiv \frac{\eta}{M_{pl}}$. In what follows we omit the bar on the dimensionless $\bar \eta$.

In an effort to go beyond the above heuristic estimates we consider the full Einstein equations and using the metric Equation (\ref{ltbmet}) and the energy-momentum tensor Equation (\ref{tmntot}) we obtain the system of \mbox {Equations \cite{BuenoSanchez:2011wr,Cho:1997rb,Sakai:1995nh}}:
\ba
	- G_0^0 &=&
      	K_2^2 (2K-3K_2^2) - 2{B'' \over A^2B} -{B'^2 \over A^2B^2}
	+2{A'B' \over A^3B} -6{B' \over A^2Br} +2{A' \over A^3r}
	-{1 \over A^2r^2} + {1 \over B^2r^2}  \nonumber\\
	&=& {8\pi \over m_{Pl}^2} \l[ \,{\dot{\Phi}^2 \over 2} + {\Phi'^2 \over 2A^2}
	+{\Phi^2 \over B^2r^2} + {\lambda \over 4}(\Phi^2 - \eta^2)^2
	+{\rho^{mat} \over 1-v^2}\, \r] \,
\label{dyneq0}
\ea
\ba
	{1 \over 2} G_{01} =
	K_2^{2\prime}  + \l( {B' \over B} + {1 \over r} \r) \l( 3K_2^2 - K \r) =
	{4\pi \over m_{Pl}^2} \l( \dot{\Phi} \Phi' -{v \over 1-v^2}\,A\,\rho^{mat} \r) \,
\label{dyneq1}
\ea
\ba
	{1 \over 2}\,(G_1^1 + G_2^2 + G_3^3 - G_0^0) &=&
      	\dot{K} - (K_1^1)^2 - 2(K_2^2)^2 \nonumber\\
	&=& {8\pi\over m_{Pl}^2} \l[ \, \dot{\Phi}^2 - {\lambda \over 4}(\Phi^2 - \eta^2)^2
	+ {1 \over 2} {1+v^2 \over 1-v^2}  \rho^{mat}\, \r] \,
\label{dyneq2}
\ea
and:
\be
	\ddot{\Phi} - K\,\dot{\Phi} - {\Phi'' \over A^2}
	- \l( -{A' \over A} + {2B' \over B} + {2 \over r} \r)\,
	{\Phi' \over A^2} + {2\Phi \over B^2 r^2} + \lambda \Phi (\Phi^2-\eta^2) = 0
\label{dyneqphi}
\ee
where prime (dot) denotes differentiation with respect to radius $r$ (time $t$) and:
\begin{equation}
K_1^1 = - {\dot{A} \over A},  K_2^2 = K_3^3 = - {\dot{B} \over B},  K = K_i^i\,
\label{kdef}
\end{equation}

The conservation of the energy momentum tensor for the matter fluid Equation (\ref{tmnmat}) in the metric Equation (\ref{ltbmet}) leads to the following equations for $v$ and $\rho^{mat}$:
\be
\frac{\dot v}{v}=(v^2-1) \frac{\dot A}{A}-\frac{v'}{A}\,
\label{vcons}
\ee
\be
\frac{\dot \rho^{mat}}{\rho^{mat}}= \frac{\dot v}{v}-2\frac{\dot B}{B} - \frac{v}{A} \frac{(\rho^{mat})'}{\rho^{mat}} -2\frac{v}{A}\frac{B'}{B}-\frac{2 v}{r \; A}
\label{rhocons}
\ee

Notice that in the homogeneous FLRW limit ($B(r,t)=A(r,t)=a(t)$, $\Phi(r,t)=\eta$) we obtain the familiar Friedman equations with $v=0$ and the usual conservation of matter equation. The system Equations (\ref{dyneq0})--(\ref{dyneqphi}), (\ref{vcons}) and (\ref{rhocons}) consists of six equations with five unknown functions $A$, $B$, $\Phi$, $v$, $\rho^{mat}$. Thus,
Equation (\ref{dyneq0}) may be considered as a constraint and the system of the other five equations may be solved with initial conditions corresponding to a homogeneous flat matter dominated universe \linebreak ($A=B=1$) with a static global monopole profile $f(r)$ ($f(0)=0$, $f(\infty)=1$) \cite{BuenoSanchez:2011wr}.
The main questions to be addresses in such a cosmological simulation are the following:
\begin{itemize}
\item
Does the monopole energy density eventually dominate over matter in the monopole core?
\item
Does the possible monopole domination eventually lead to accelerating expansion in the \mbox {monopole core?}
\item
Can this cosmological expansion in the core fit the cosmological data?
\end{itemize}

\begin{figure}[H]
\centering
\epsfig{file=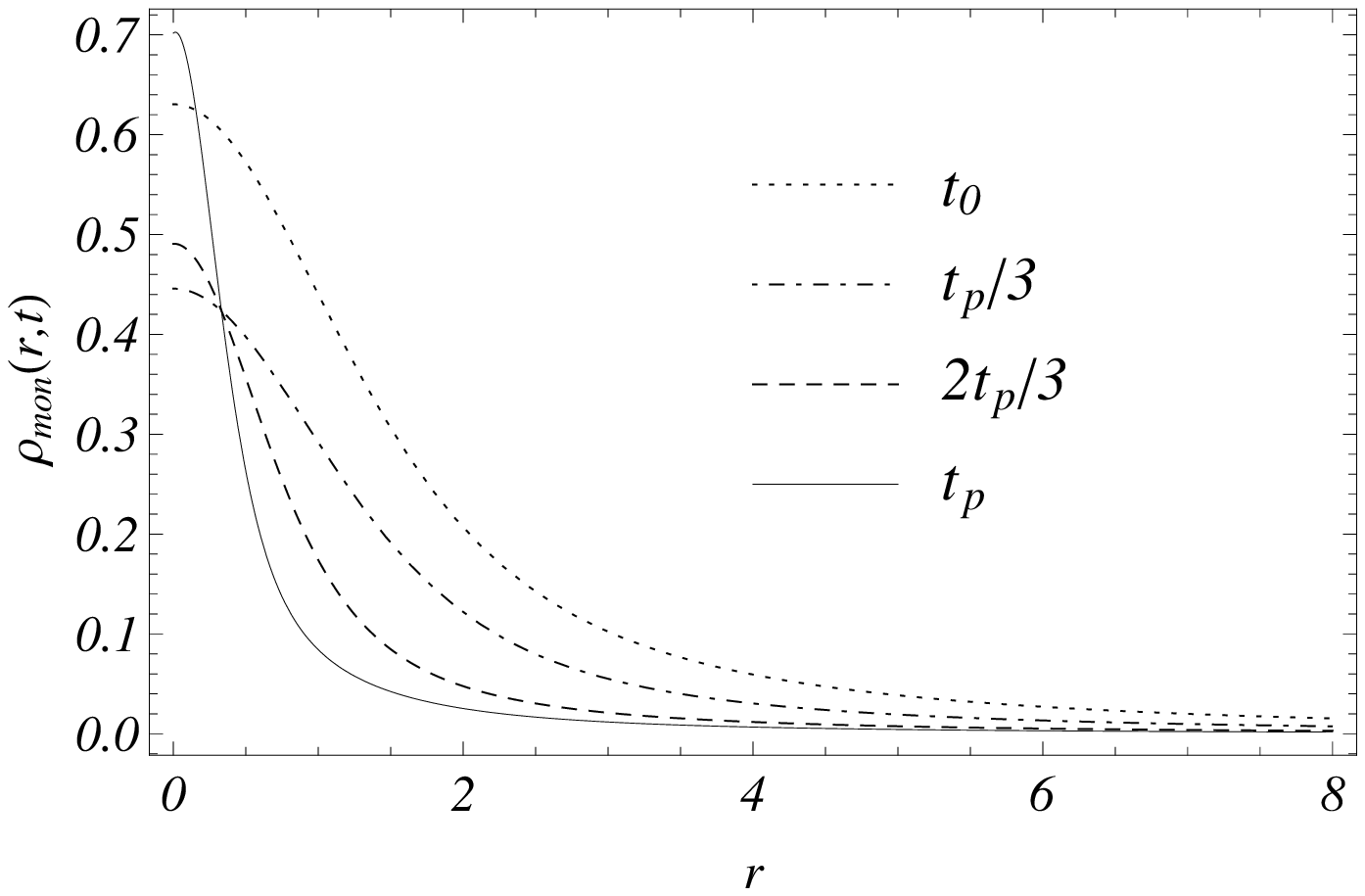,width=7cm}
\epsfig{file=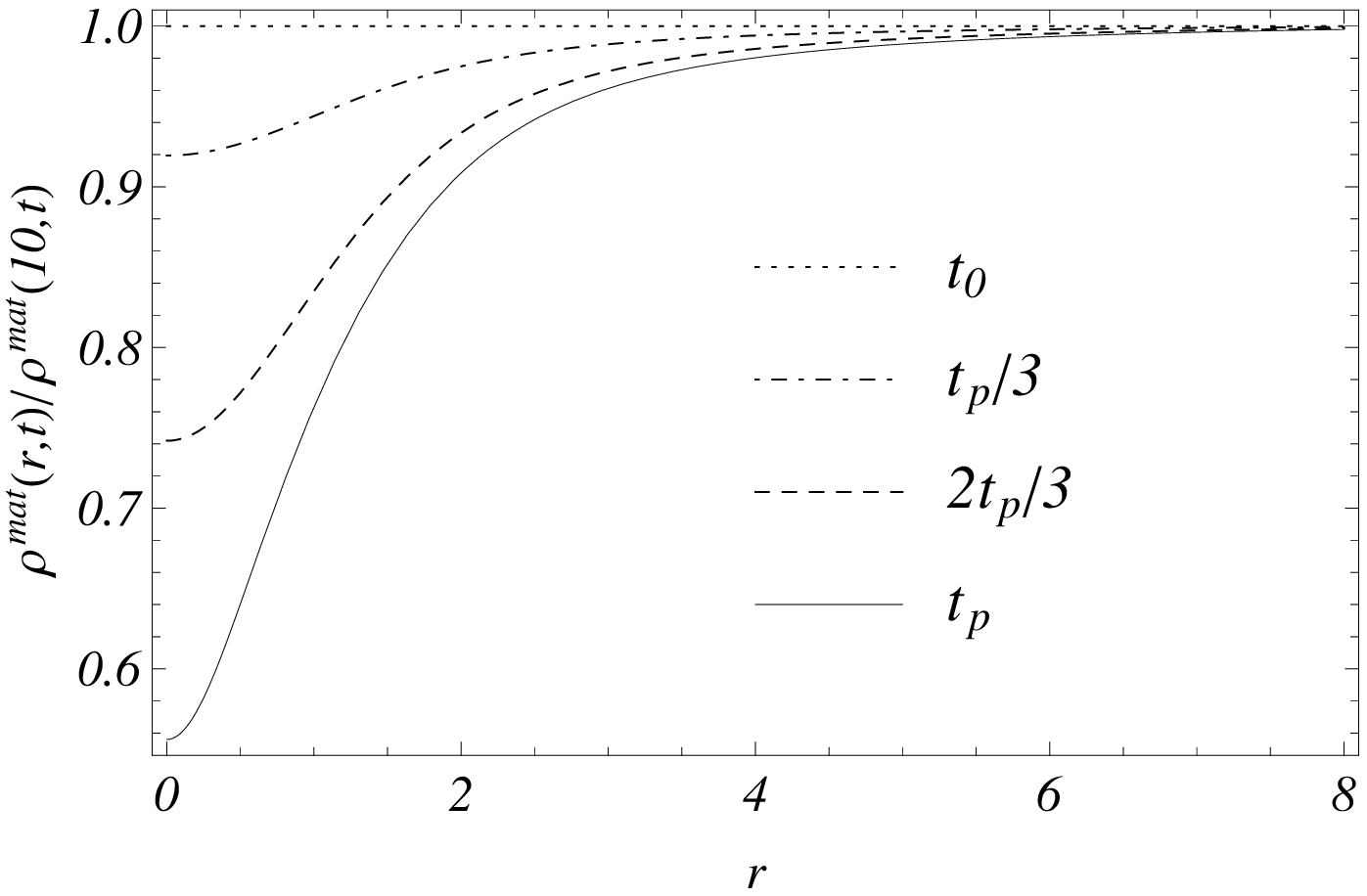,width=7cm}
\caption{Evolution of the dark energy (left panel) and properly normalized matter (right panel) density profiles.The profiles shown correspond to the initial time $t_0$ (dotted line), $t_p/3$ (dot-dashed line), $2t_p/3$ (dashed line), $t_p$ (solid line, present time). Notice the matter underdensity that develops in the monopole core while the monopole density appears to slowly collapse to the center in comoving coordinates. The radial unit is the monopole core size $\delta$ (from   \cite{BuenoSanchez:2011wr}). }\label{densevol}
\end{figure}

The answer to all three of the above questions turns out to be positive. The evolution of the densities is shown in Figure \ref{densevol}. As expected, matter develops an underdensity in the core due to the repulsive gravity of the vacuum energy.

\begin{figure}[H]
\centering
\epsfig{file=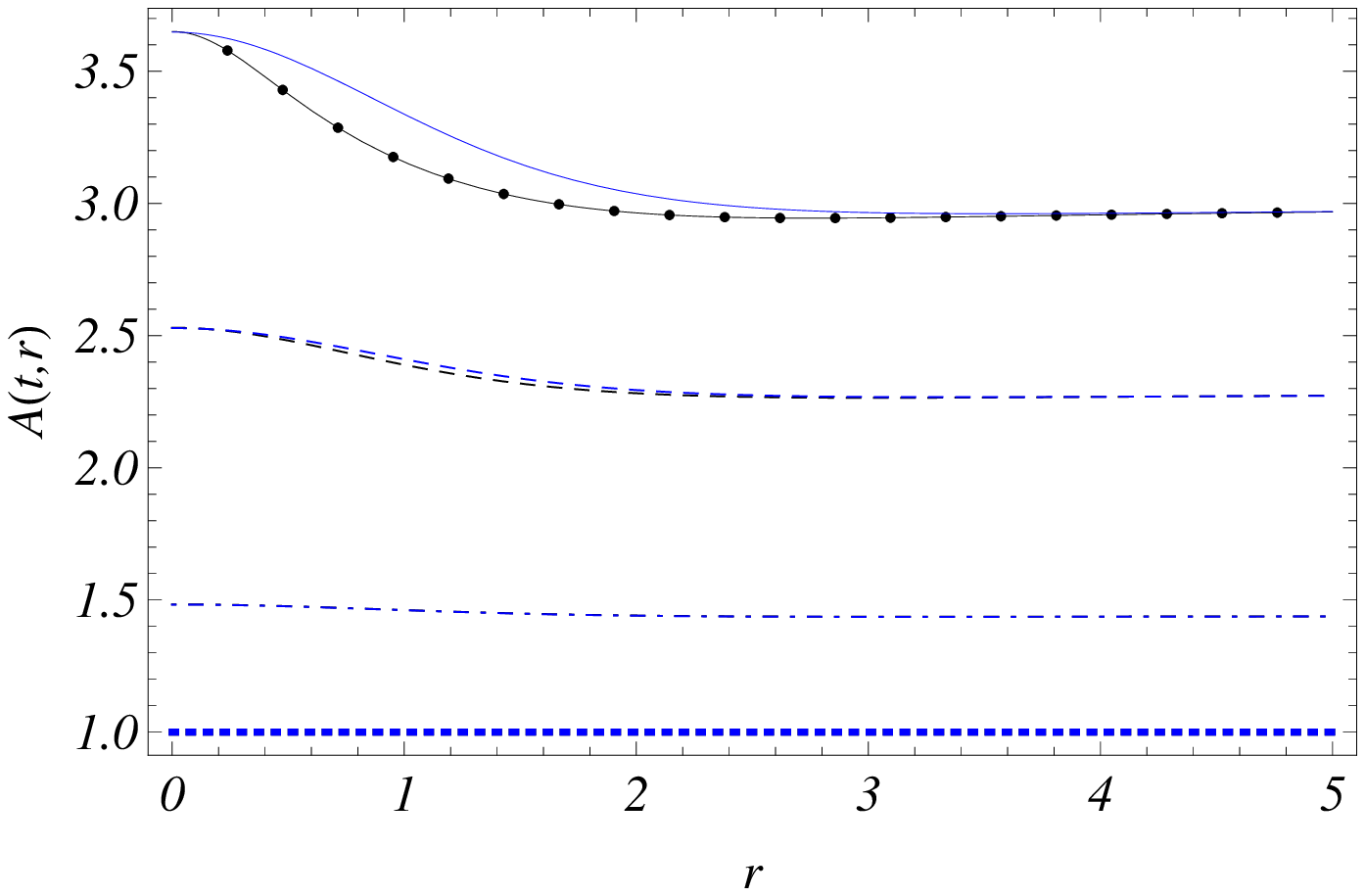,width=7.5cm}
\epsfig{file=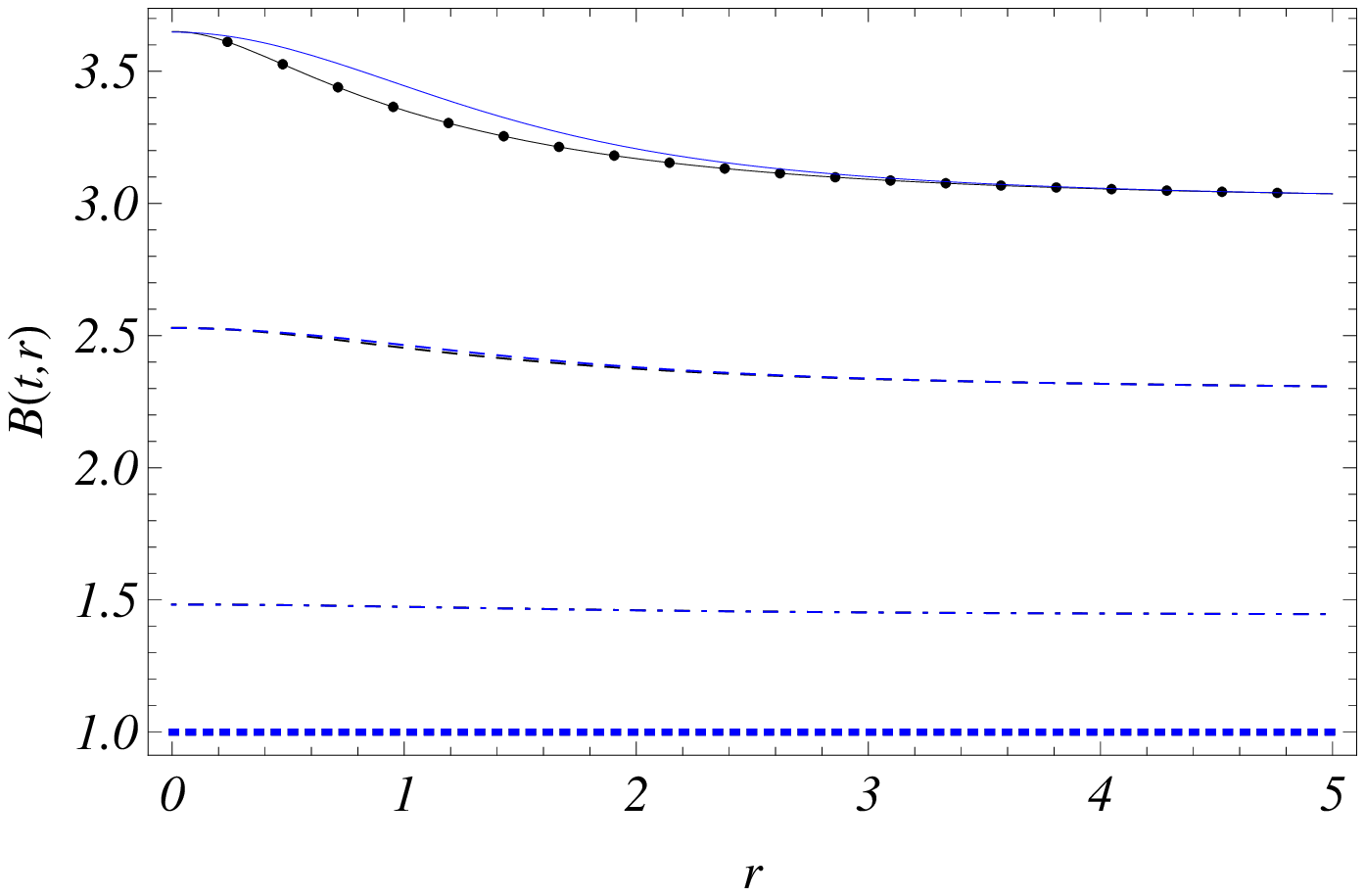,width=7.5cm}
\caption{Evolution of the scale factors $A(r,t)$ and $B(r,t)$ profile for $\eta=0.1$ (black lines) and $\eta=0.6$ (blue lines) from the time $t_0$ up to the present time $t_p$. A mesh of dots is superimposed to the curve corresponding to $\eta=0.1$, $t=t_p$ to distinguish it from the curve obtained for $\eta=0.6$, $t=t_p$. The profiles shown correspond to $t_0$ (dotted line), $t_p/3$ (dot-dashed line), $2t_p/3$ (dashed line), $t_p$ (solid line). Higher curves correspond to more recent times.  The (rescaled) comoving region of accelerated expansion is slightly smaller in the case of $\eta<0.3$ since the monopole core does not expand in this case (from  \cite{BuenoSanchez:2011wr}). }\label{scalfacev}
\end{figure}

Clearly the vacuum energy eventually dominates in the monopole core. The time evolution of the scale factors $A$ and $B$ is shown in Figure \ref{scalfacev} for two values of $\bar \eta$. Clearly the core experiences a faster (accelerating) expansion compared to regions outside the monopole core. For ${\bar \eta} > 0.3 $ the comoving monopole core size appears to expand in agreement with previous studies of topological inflation.

\begin{figure}[H]
\centering
\epsfig{file=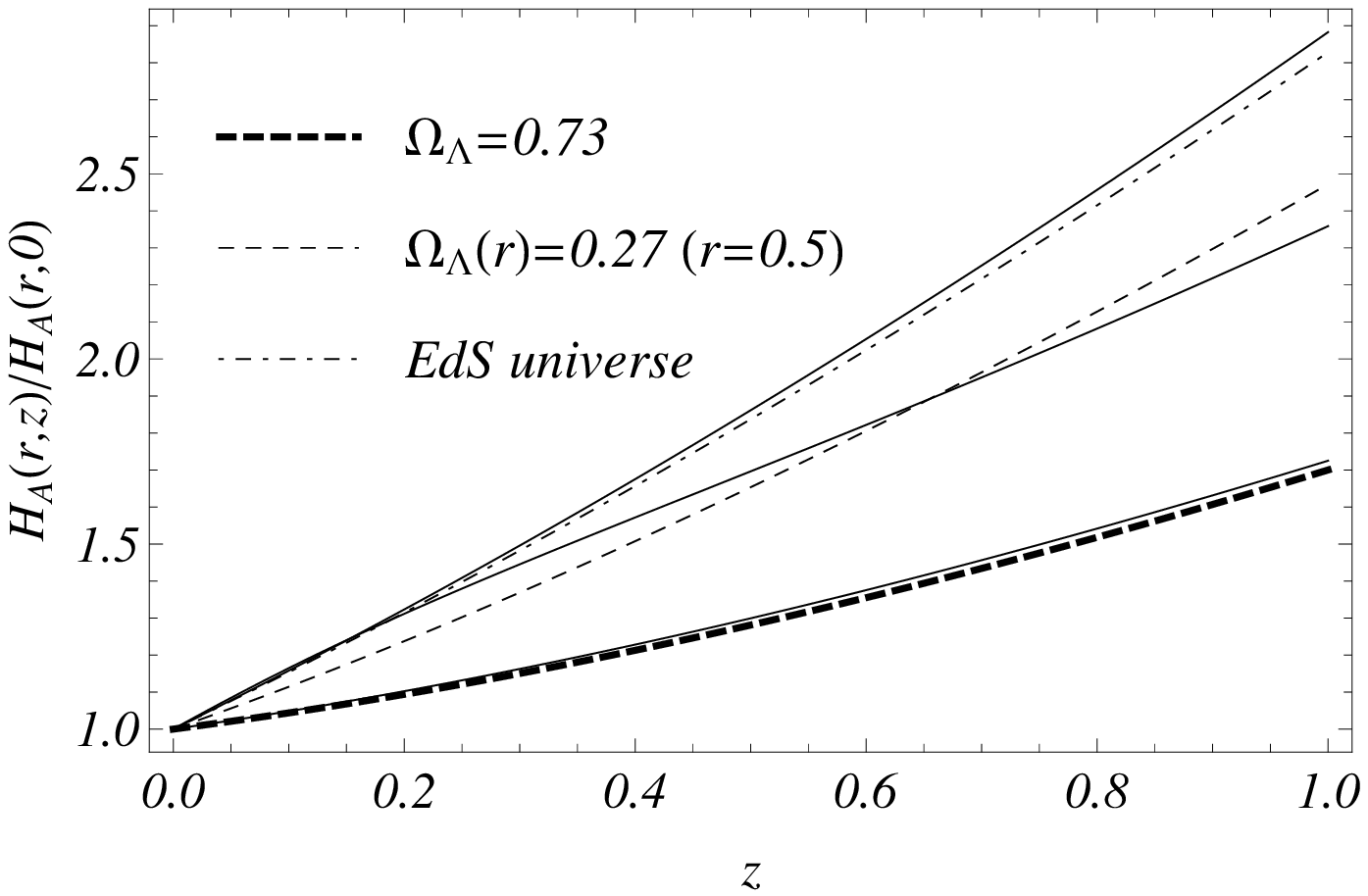,width=7.5cm}
\epsfig{file=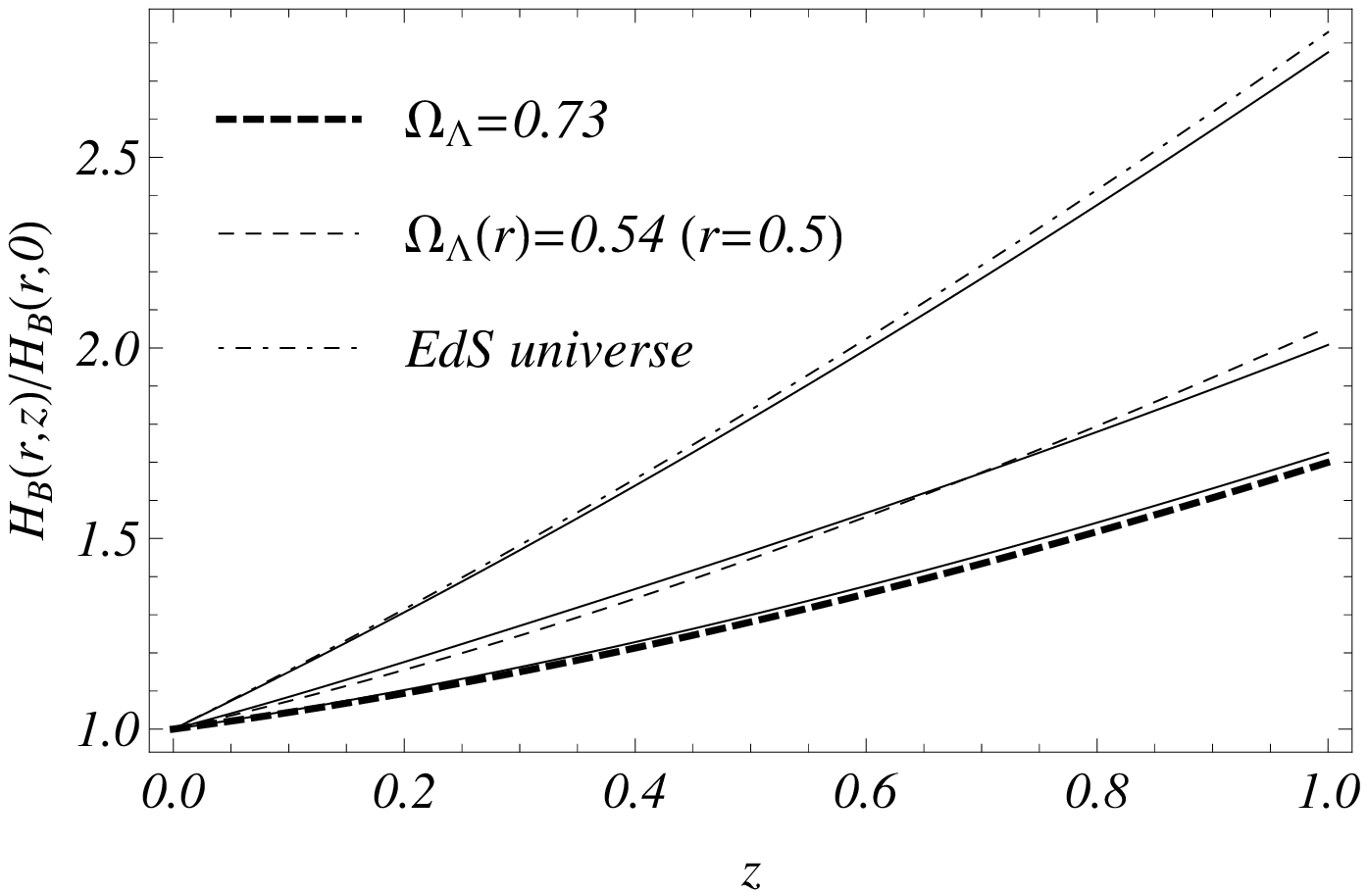,width=7.5cm}\caption{Ratios $H_A(r,z)/H_A(r,0)$ and $H_B(r,z)/H_B(r,0)$ for $r=0$ (lower solid line), $r=0.5$ and $r=5$ (upper solid lines) along with the corresponding $H_\Lambda(r,z)/H_0$ for \mbox {$\Omega_\Lambda=0.73$} (thick, dashed line) and for $\Omega_\Lambda=0$ (dot-dashed line). Also included for comparison $H_\Lambda(r,z)/H_0$ for the best fit values of $\Omega_\Lambda(r)$: $\Omega_\Lambda(0.5)\simeq0.27$ (left panel) and $\Omega_\Lambda(0.5)\simeq0.54$ (right panel). We have set $\eta=0.1$  (from  \cite{BuenoSanchez:2011wr}).}\label{hubevol}
\end{figure}

In Figure \ref{hubevol} we show the evolution of the Hubble expansion rate ratios $H_A(r,z)/H_A(r,0)$ and $H_B(r,z)/H_B(r,0)$ at various comoving distances from the monopole center. Superposed is shown the {\lcdm} Hubble expansion rate for the best fit values of $\Omega_\Lambda$. We have chosen the present time such that the best fit value at the core center is $\Omega_\Lambda = 0.73$. Clearly the expansion rates at all distances are almost identical with {\lcdm} with diminishing choices of $\Omega_\Lambda$. Thus, for a core size larger than a few Gpc we anticipate good agreement of the predictions of the model with SnIa and other cosmological data.

\vspace{-12pt}

\subsection{Spherical Dark Energy Overdensity}

Even though topological quintessence is a well defined physical model based on specific dynamics, the predicted nonlinear evolution of dark energy complicates the detailed comparison with cosmological observation. This \scalebox{.95}{comparison, is significantly simplified if we approximate the dark energy inhomogeneity} as a static spherically symmetric fluid with negative pressure in a flat background. In particular we may consider the energy momentum tensor:
\be
T_\nu^\mu = \diag(-\rho(r)-\rho_M(r,t),p_r(r),p_t(r),p_t(r))
\label{enmom}
\ee
where $\rho_M(r,t)$ is the matter density and we have allowed for a general inhomogeneous and spherically symmetric static fluid with:
\be
(T_f)_\nu^\mu = \diag(-\rho(r),p_r(r),p_t(r),p_t(r))
\label{enmommon}
\ee
(where $p_t(r)$ is the transverse pressure). This expression is motivated by the energy-momentum of a global monopole, which asymptotically is of the form~\cite{Barriola:1989hx}:
\be (T_{\rm mon})_\nu^\mu = \diag(-\eta/r^2,-\eta/r^2,0,0) \label{tmnmonas} \ee
with $\eta$ the symmetry-breaking scale related to the formation of the monopole.

\begin{figure}[H]
\centering
\includegraphics[trim = 0mm 230mm 110mm 12mm, clip=true, width=0.8\textwidth]{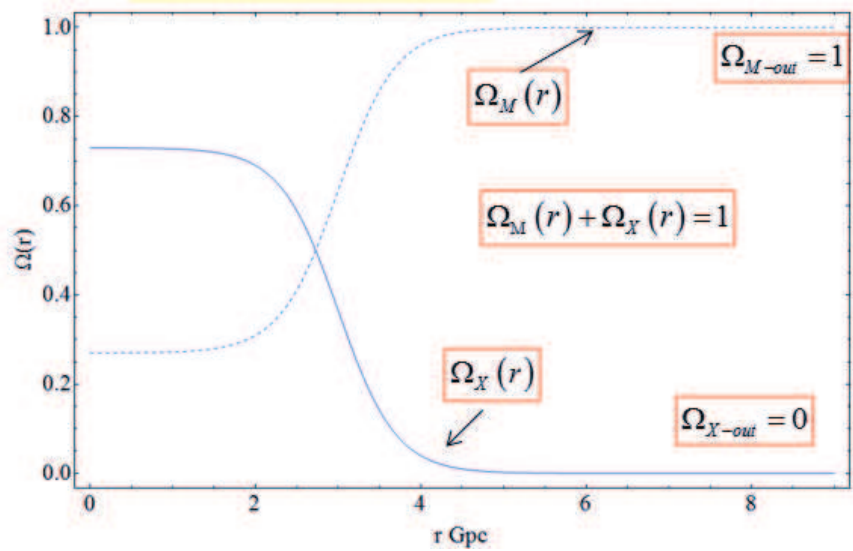}
%\rotatebox{0}{\hspace{0cm}\resizebox{1.7 \textwidth}{!}{\includegraphics{profiles.eps}}}
\hspace{0pt}
\caption{The assumed matter and dark energy profiles are consistent with flatness and with matter domination outside the inhomogeneity. Such profiles would develop dynamically in the context of topological quintessence (see Figure \ref{densevol}). \vspace{-6pt}  \label{profiles}}
\end{figure}

Using Einstein equations, it may be shown \cite{Grande:2011hm} that the metric corresponding to the diagonal energy momentum tensor Equation (\ref{enmom}) is of the LTB form Equation (\ref{ltbmet2}). Using Einstein equation we obtain the cosmological Equation (\ref{freq3}). In contrast to the matter void here we set $\Omega_c(r)=0$ and use the profiles Equations (\ref{omegam}) and (\ref{omegax}) for the matter and dark energy fluids which are consistent with flatness ($\Omega_{M,\text{out}}=1-\Omega_{X,\text{out}}=1,  $\mbox { Figure \ref{profiles})}.

\begin{figure}[H]
\centering
\includegraphics[trim = 0mm 190mm 100mm 12mm, clip=true, width=0.58\textwidth]{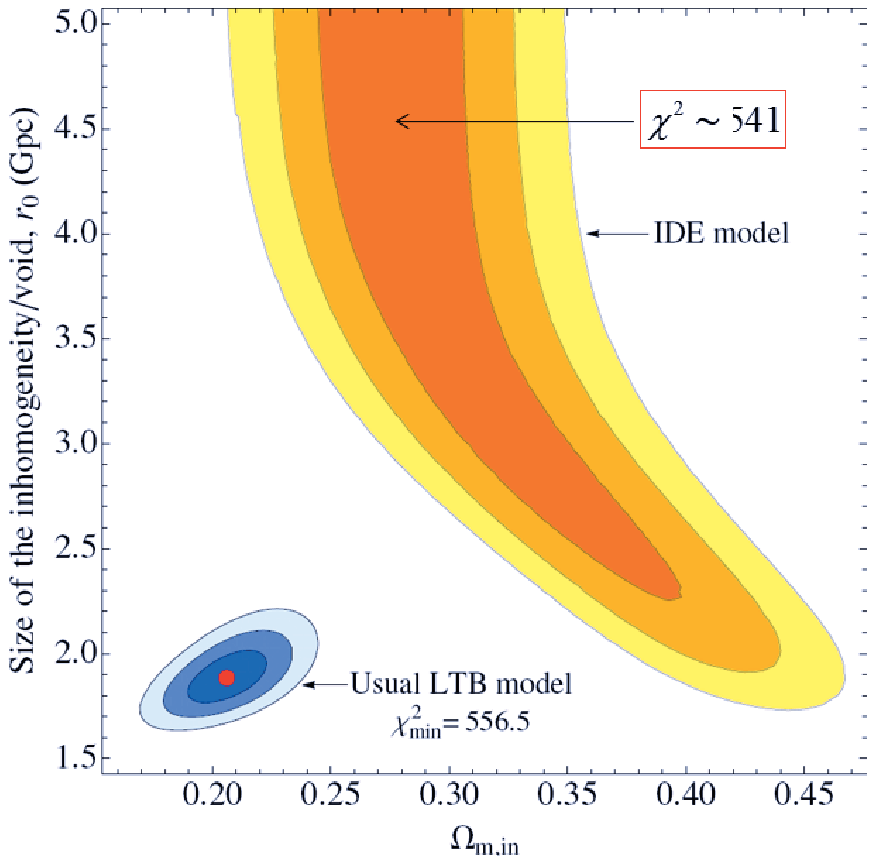}
%\rotatebox{0}{\hspace{0cm}\resizebox{1.7 \textwidth}{!}{\includegraphics{ideconstr.eps}}}
\hspace{0pt}
\caption{The $1$$\sigma$ and $2$$\sigma$ parameter contours for the inhomogeneous dark energy (IDE) fluid model obtained using the Union 2 data, compared with the corresponding contours of the usual LTM matter void model. Notice that for an inhomogeneity size larger than the range of the dataset (about 4 Gpc), the model reduces to \lcdm ($\Omega_m=2.7$, $\Omega_X=0.73$. The transition scale was fixed to $\Delta r=$ 0.35 Gpc (from  \cite{Grande:2011hm}). \vspace{-6pt} \label{ideconstr} }
\end{figure}

\vspace{-12pt}
Considering an observer at the center of the inhomogeneity, we may use the geodesics \cite{Grande:2011hm,Alnes:2006uk} to obtain the luminosity distance as described above Equation (\ref{lumidi}). By comparing the predicted luminosity distance with the Union 2 data \cite{Amanullah:2010vv}, we obtain \cite{Grande:2011hm} constraints on the model parameters $\Omega_{M,\text{in}}=1-\Omega_{X,\text{in}}$ and $r_0$ (the size of the inhomogeneity) using the maximum likelihood \mbox {method \cite{Grande:2011hm}}. The allowed range of parameters is significantly larger for the inhomogeneous dark energy model (IDE) and as expected, the model reduces to {\lcdm} for a scale of inhomogeneity that is larger than the range of the \linebreak Union 2 dataset (see Figure \ref{ideconstr}).

\begin{figure}[H]
\centering
\includegraphics[trim = 0mm 190mm 100mm 12mm, clip=true,width=0.8\textwidth]{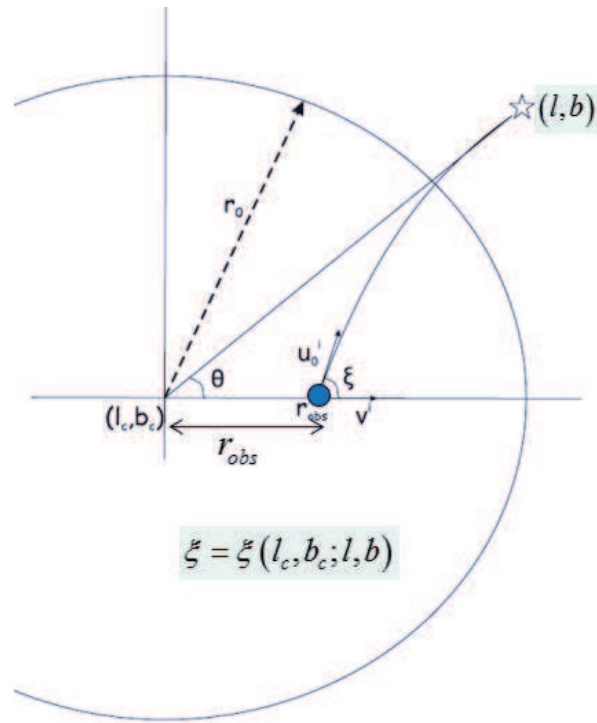}
%\rotatebox{0}{\hspace{0cm}\resizebox{0.39 \textwidth}{!}{\includegraphics{offcentergeom.eps}}}
\hspace{0pt}
\caption{The lightlike geodesics for an off-center depend on two additional parameters $r_{obs}$ and $\xi$. \vspace{-6pt}\label{offcentergeom}}
\end{figure}

It is straightforward to extend this analysis to the case of an off-center observer \cite{Grande:2011hm,Balcerzak:2013afa}. In this case the light-like geodesics of the photons coming from the SnIa, depend on two additional parameters: the distance of the observer from the center of the inhomogeneity ($r_{obs}$) and the incidence angle $\xi$ of the photons with respect to the displacement axis (Figure \ref{offcentergeom}). The angle $\xi$ in turn may be expressed in terms of the direction of the center in galactic coordinates $(l_c,b_c)$ and the direction the supernova that emits the photon $(l,b)$ ($\xi=\xi(l_c,b_c,l,b)$). Thus, it is straightforward \cite{Grande:2011hm} to obtain an expression of the predicted luminosity distance as $d_L(z,r_0,\Omega_{X,\text{in}};r_{obs},\xi)$ and use the Union 2 data (with the directions $(l,b)$ of each SnIa) with the maximum likelihood method to derive $\chi^2(r_0,\Omega_{X,\text{in}};r_{obs},l_c,b_c)$.

\vspace{-18pt}
\vspace{-6pt}

\begin{figure}[H]
\centering
\includegraphics[trim = 5mm 235mm 110mm 15mm, clip=true, width=\textwidth]{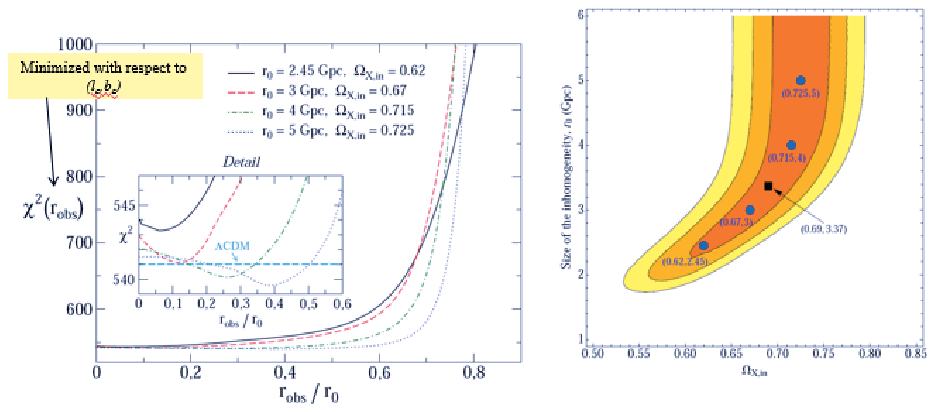}
%\rotatebox{0}{\hspace{0cm}\resizebox{1.7 \textwidth}{!}{\includegraphics{chi2robs.eps}}}
\hspace{0pt}
\caption{The dependence of $\chi^2$ on the displacement $r_{obs}$ of the observer (left panel). Minimization with respect to the direction of the inhomogeneity center has been performed. We have considered four pairs ($r_0, \Omega_{X,\text{in}}$) that provide good fits for the on center observer (right panel). Notice that improved fit with respect to {\lcdm} (horizontal blue dashed line) is obtained only for large scale inhomogeneities ($r_0>3.5$ Gpc) (from \cite{Grande:2011hm}).  \label{chi2robs} }
\end{figure}

In Figure \ref{chi2robs} we show the dependence of $\chi^2$ on the displacement $r_{obs}$ of the observer (left panel). Minimization with respect to the direction of the inhomogeneity center has been performed. We have considered four pairs ($r_0, \Omega_{X,\text{in}}$) that provide good fits for the on center observer (right panel). Notice that improved fit with respect to {\lcdm} (horizontal blue dashed line) is obtained only for large scale inhomogeneities ($r_0>3.5$ Gpc).

\vspace{-48pt}

\begin{figure}[H]
\centering
\includegraphics[trim = 0mm 0mm 0mm 0mm, clip=true, width=0.75\textwidth]{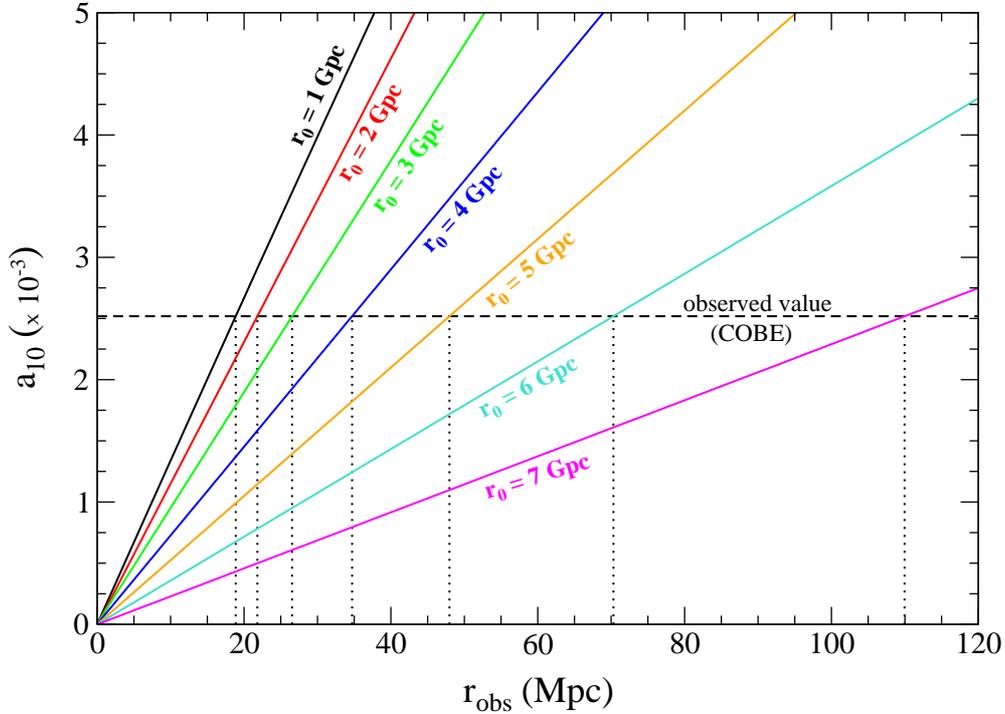}
%\rotatebox{0}{\hspace{0cm}\resizebox{0.7\textwidth}{!}{\includegraphics{cmbdip.eps}}}
\hspace{0pt}
\caption{The predicted CMB dipole moment $a_{10}(r_{obs})$ for $\Omega_{X,\text{in}}=0.69$ and various values of the inhomogeneity scale $r_0$. The dashed line corresponds to the observed value (whose uncertainty is too small to be shown (from   \cite{Grande:2011hm}).\label{cmbdip} }
\end{figure}

It is straightforward \cite{Grande:2011hm}  to also use geodesics from the last scattering surface to the present time to derive the CMB photon geodesics for an off center observer. From these we may obtain the last scattering surface redshift $z_{ls}$ as a function of the inhomogeneity parameters and in particular $\xi$ and $r_{obs}$ (assuming fixed the inhomogeneity parameters $r_0, \Omega_{X,\text{in}}$). After obtaining $z_{ls}(\xi,r_{obs})$ we find the angular dependence of the temperature as:
\be T(\xi,r_{obs})=\frac{T_{ls}}{1+z_{ls}(\xi,r_{obs})} \label{tempprof}
\ee
where $T_{ls}$ is the observed temperature of the last scattering surface. This is assumed to be uniform since we are focusing on the {\it additional} temperature fluctuations induced by the displacement of the observer. We then convert it to temperature fluctuations from the mean and expand in spherical harmonics to find the multipole coefficients.
Comparison with the measured values of the moments (particularly of the dipole) leads to constraints on $r_{obs}$.
It is straightforward to obtain numerically  $a_{10}(r_{obs})$, the additional CMB fluctuations dipole moment induced by the shift $r_{obs}$ of the observer. This quantity is shown in Figure \ref{cmbdip} for various values of the inhomogeneity scale $r_0$. In the same plot we show the value of the measured CMB dipole (dashed line)~\cite{Lineweaver:1996qw}:
\begin{eqnarray}
\label{measdip}
\left(\frac{\Delta T}{\bar T}\right)_{10}&=&\frac{3.35 \times {10^{-3}}\,{\rm K}}{2.725\,{\rm K}}=(1.230\pm0.013)\times 10^{-3}\nonumber\\
a_{10}&=&\sqrt{\frac{4\pi}{3}}\left(\frac{\Delta T}{\bar T}\right)_{10}\simeq(2.52\pm0.03)\times 10^{-3}\nonumber\\[-14mm]
&&\label{cobedip2}
\end{eqnarray}

\vspace{18pt}

Clearly, for $r_0<5{\rm \,}$Gpc the observer is confined to be in a sphere which is a fine tuned small spatial fraction of the dark energy inhomogeneity ($f(r_0)\equiv(\frac{r_{obs-max}}{r_0})^3\simeq 10^{-6}$) which implies severe fine tuning of the model as in the case of LTB models with matter. This is shown in Figure~\ref{fracvol} where $f(r_0)$ is plotted for values of $r_0$ between 1 and 7 Gpc. As shown in that figure, the fine tuning starts to reduce ($f(r_0)$ increases)  as the size of the inhomogeneity increases beyond $6{\rm \,}$Gpc. We anticipate it to disappear as the size of the inhomogeneity reaches $\sim$$14$ Gpc (the comoving distance to the last scattering surface).

\begin{figure}[H]
\centering
\includegraphics[trim = 0mm 190mm 70mm 12mm, clip=true,width=0.9\textwidth]{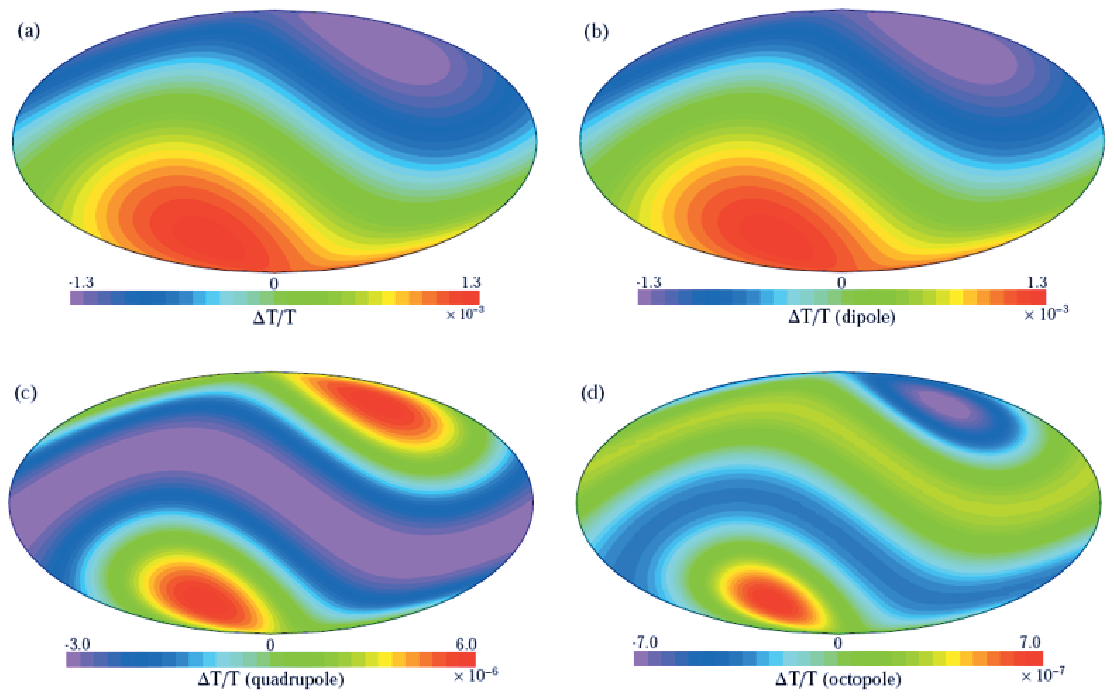}
%\rotatebox{0}{\hspace{0cm}\resizebox{1.1 \textwidth}{!}{\includegraphics{cmbmap.eps}}}
\vspace{-24pt}
\caption{The predicted CMB maps of additional temperature fluctuations in galactic coordinates corresponding to the full map ({\bf a}); dipole ({\bf b}) quadrupole ({\bf c}) and octopole ({\bf d}). We have selected the direction of the preferred axis to coincide with the direction of the observed CMB dipole and used $(r_0,\Omega_{X, \text{in}})=(3.37 {\rm \,Gpc},0.69)$, $r_{obs}=30$ Mpc (from   \cite{Grande:2011hm}). \vspace{-18pt} \label{cmbmaps}}
\end{figure}

The predicted CMB maps of additional temperature fluctuations in galactic coordinates corresponding to the full map (a), dipole (b) quadrupole (c) and octopole (d) are shown in Figure  \ref{cmbmaps}. Notice the rapid decrease of the magnitude of moments higher than the dipole. Even though the order of magnitude of the  dipole and quadrupole is comparable with observations, higher moments have negligible magnitude. This is anticipated since the displacement of the observer is expected to modify the CMB fluctuations only at the largest scales.

\begin{figure}[H]
\centering

\includegraphics[trim = 0mm 0mm 0mm 0mm, clip=true, width=0.63\textwidth]{fracvol.eps}
%\rotatebox{0}{\hspace{0cm}\resizebox{0.7\textwidth}{!}{\includegraphics{fracvol.eps}}}

\caption{The spatial fraction $f(r_0)\equiv(\frac{r_{obs-max}}{r_0})^3$ where the observer needs to be \protect\linebreak confined in order to be consistent with the value of the observed CMB dipole, for $(r_0,\Omega_{X, \text{in}})=(3.37 {\rm \,Gpc},0.69)$ (from \cite{Grande:2011hm}). \label{fracvol}}
\end{figure}

\vspace{-8pt}

\section{Conclusions}

It has become evident during the last few years that early hints for deviation from the cosmological principle and statistical isotropy are being accumulated.  This appears to be one of the most likely directions which may lead to new fundamental physics in the coming years. We have reviewed these hints, including the large scale CMB anomalies (CMB asymmetries), the large scale peculiar velocity flows on scales of  100 Mpc  and up to a  Gpc  (dark flow), the preliminary evidence for the spatial variation of the fine structure constant along a preferred axis on  Gpc scales, the coherent variation of quasar optical polarization on  Gpc scales, the existence of the Large Quasar Structure with a Gpc scale and the mild hints for anisotropic expansion rate. We have pointed out the fact that most of the above cosmological observations are associated with the existence of specific cosmological directions which appear to be abnormally aligned with each other.

The Gpc scale associated with most of these effects indicates that their physical origin has appeared at late cosmological times. This possibility is further amplified by the fact that the CMB power asymmetry is significantly diminished if the ISW effect is subtracted from the CMB maps \cite{Rassat:2013gla}. It is therefore natural to attempt to associate the physical origin of these effects with large scale dark energy or matter inhomogeneities. Hubble scale matter inhomogeneities in the form of voids are severely constrained by cosmological observations and are practically ruled out. On the other hand, dark energy Hubble scale inhomogeneities are observationally viable and reduce to standard {\lcdm} when the inhomogeneity scale exceeds the horizon scale. A simple mechanism that can give rise to a cosmological preferred axis is based on an off-center observer in a spherical dark energy inhomogeneity. We have shown that this mechanism can naturally give rise to the observed alignment of many of the above described \mbox {cosmic anisotropies.}

The repulsive gravity properties of dark energy and its high sound velocity imply that any large scale dark energy inhomogeneity would tend to smooth out. This is a challenge for the above described mechanism. Nontrivial topology however can lead to sustainable Hubble scale dark energy inhomogeneities. If we happen to leave in the core of a Hubble scale recently formed topological defect we would naturally experience Hubble scale dark energy inhomogeneities and the existence of a preferred axis due to our displacement from the defect center. Therefore, topological quintessence constitutes a physical mechanism to produce Hubble scale dark energy inhomogeneities. We have shown that such a mechanism can also give rise to a Dark Energy Dipole,  Large Scale Velocity Flows, Fine Structure Constant Dipole and a CMB Temperature Asymmetry aligned with each other.  Other effects which may also occur (quasar polarization alignment etc) constitute interesting potential extensions of this research direction.

 %\lcdm predicts significantly smaller amplitudeough and scale of flows than what observations indicate. It has been found that the dipole moment (bulk flow) of a combined  peculiar velocity sample extends \cite{Watkins:2008hf} on scales up to $100 h^{-1}Mpc$ ($z\leq 0.03$) with amplitude larger than $400 km/sec$. The direction of the flow has been found consistently to be approximately in the direction $l \simeq 282^\circ$, $b\simeq 6^\circ$. Other independent studies have also found large bulk velocity flows on similar directions \cite{Lavaux:2008th} on scales of about $100 h^{-1}Mpc$ or larger \cite{Kashlinsky:2008ut}. The expected $rms$ bulk flow in the context of \lcdm normalized with WMAP5 $(\omm,\sigma_8)=(0.258,0.796)$ on scales larger than $50h^{-1}Mpc$ is approximately $110 km/sec$. The probability that a flow of magnitude larger than $400km/sec$ is realized in the context of the above \lcdm normalization (on scales larger than $50h^{-1} Mpc$) is less than $1\%$. A possible connection of such large scale velocity flows and cosmic acceleration may be found in Ref. \cite{Tsagas:2009nh}.

\acknowledgements{Acknowledgments}

This research has been co-financed by the European Union (European
Social Fund--ESF) and Greek national funds through the Operational
Program ``Education and Lifelong Learning'' of the National Strategic
Reference Framework (NSRF)--Research Funding Program: ARISTEIA.\@
Investing in the society of knowledge through the European Social
Fund.

%%%%%%%%%%%%%%%%%%%%%%%%%%%%%%%%%%%%%%%%%%

\conflictofinterests{Conflicts of Interest}

The author declares no conflict of interest.

%=================================================================
% References: Variant A
%=================================================================
% Back Matter (References and Notes)
%----------------------------------------------------------
% Style and layout of the references
\bibliographystyle{mdpi}
\makeatletter
\renewcommand\@biblabel[1]{#1. }
\makeatother

%=================================================================
% References:  Variant B
%=================================================================
% Use the following option to include external BibTeX files:
%\bibliography{lite}
%\bibliographystyle{mdpi}

\end{document}